\documentclass[superscriptaddress,floatfix,aps,pre,amsmath,amssymb,twocolumn,shortbibliography]{revtex4-2}
\usepackage[utf8]{inputenc}
\usepackage[T1]{fontenc}
\usepackage{lmodern}
\usepackage{graphicx}
\usepackage{bm}
\usepackage[english]{babel}
\usepackage{microtype}
\usepackage{hyperref}
\hypersetup{colorlinks,allcolors=black}
\begin{document}
\title{An approximate separation of the electron and positron degrees of freedom in the Dirac-Heisenberg-Wigner formalism}
\author{Gert Brodin}
\email{gert.brodin@umu.se}
\affiliation{Department of Physics, Ume{\aa} University, SE--901 87 Ume{\aa}, Sweden}
\author{Sebastian Lundstr\"om}
\email{sebastian.lundstroem@ipp.mpg.de}
\affiliation{Department of Physics, Ume{\aa} University, SE--901 87 Ume{\aa}, Sweden}
\begin{abstract}
Here we study the quantum relativistic phase-space description of the Dirac field referred to
as the Dirac--Heisenberg--Wigner (DHW) formalism. By applying a systematic expansion
procedure in weak electromagnetic fields (compared to the Schwinger critical field) and slow
temporal and long spatial scales (compared to the Compton scales), the 16 scalar phase-space
functions of the DHW equations can be reduced to 8 scalars. By making a
further change of variables, the equations can be interpreted as four evolution
equations for the electron degrees of freedom and four for the positron degrees of freedom. We refer to these equations as the weakly coupled electron-positron (WCEP) limit of the DHW equations. Unlike the case after a Foldy--Wouthuysen transformation has been made, the electron and positron degrees of freedom remain weakly coupled. Naturally, further approximations can
be made on the 4+4 equations so that only 4 equations for one particle species remain. Importantly, the
expansion scheme can be carried out to all orders, in which case the reduced 4+4 model, contrary to other reduced models, will cover most quantum relativistic effects. Specifically, this includes vacuum polarization, a phenomenon that otherwise has been added phenomenologically, but here appears naturally as a feature of the general expansion scheme.  
\end{abstract}
\maketitle
\section{Introduction}
\label{sec:introduction}

Quantum kinetic theory extends the classical phase-space description of
plasmas to regimes where the wave nature of particles, Fermi
statistics, and spin properties influence the dynamics. Such regimes occur in dense
plasmas and condensed matter systems. But quantum effects can also be
important when electromagnetic fields or their gradients are sufficiently
strong. The appropriate description depends on which effects must be
kept and on the relevant spatial and temporal scales, and the electromagnetic field strengths.
Reviews of various kinetic models are given by e.g. 
Refs. \cite{melrose2020quantum,brodin2022quantum,manfredi2019phase}. Similar considerations for fluid modes are given, for example, in Ref. \cite{manfredi2021fluid}. More general quantum kinetic approaches
also incorporate exchange and correlation effects beyond the mean-field
approximation~\cite{bonitz2016quantum}. In the present paper, we consider relativistic
quantum plasmas within a collisionless mean-field description, with
particular emphasis on the relation between particle transport, spin dynamics,
and the response of the Dirac vacuum.

For non-relativistic electrons, a Wigner transformation of the
Schr\"odinger equation provides a kinetic description of (quantum) particle
dispersion. Starting instead from the Pauli equation introduces spin
precession and the magnetic dipole force, together with the magnetization current. These effects can be described using coupled scalar and
vector distribution functions, as in the semi-classical model of Hurst
\textit{et al.}~\cite{Hurst2014}. Quantum hydrodynamic models provide an
alternative description when a fluid closure can be made; examples are hydrodynamic models put forward in
Refs. \cite{asenjo2011hydrodynamical,andreev2015separated,haas2010fluid,misra2010generation}. The kinetic treatment keeps the
momentum dependence needed to describe resonant
wave--particle interactions in spin-polarized
plasmas~\cite{zamanian2010scalar, brodin2008effects,bakri2026wave,del2017spin}.

Relativistic extensions introduce further couplings between spin dynamics and
orbital motion. In the weakly relativistic regime, spin--orbit
interaction, Thomas precession, and spin-dependent polarization enter
the kinetic description~\cite{asenjo2012semi,Hurst2017}. Fully
relativistic spin kinetics has also been developed, i.e., without expanding in
the particle momentum~\cite{ekman2017relativistic}. In that particular formulation,
a spin transform introduces spin as an additional independent variable,
whereas scalar and vector formulations encode the spin dependence in a
set of phase-space functions. 

In general, variables used in quantum relativistic formulations must be distinguished carefully: a spin-like vector constructed from
Dirac bilinears does not necessarily coincide with the spin distribution in
the particle rest frame. The relativistic precession of the latter is
governed by the Bargmann--Michel--Telegdi (BMT)
equation~\cite{BMT1959}.

A broad description follows by applying a gauge-invariant Wigner
transformation directly to the Dirac theory. Covariant quantum transport
equations were developed by Vasak \textit{et al.}~\cite{vasak1987quantum},
and the Dirac--Heisenberg--Wigner (DHW) formalism with a similar scope was formulated by Bialynicki-Birula \textit{et al.}~\cite{BialynickiBirula1991}. While the DHW formalism is not covariant, it has the advantage of avoiding extra constraint equations (in addition to the evolution equations), which are part of the covariant formulation. The usefulness of the DHW theory for problems involving electron-positron pair production (through the Schwinger mechanism and by high-frequency fields \cite{li2021enhanced,hebenstreit2011pair,aleksandrov2020pair,edwards2025relativistic}), for self-consistent plasma dynamics \cite{al2021plasma,brodin2023plasma,nematizadeh2021study}, and also for applications to interband dynamics in graphene materials \cite{morandi2011wigner,akal2016low} has been well established in the literature.

The full DHW description is, however, more extensive than what is needed
when the fields are weak compared with the Schwinger critical fields (electric field $E_c\approx1.3\times10^{18}$ V/m and magnetic field $B_c\approx4.4\times10^9$ T)
and vary slowly on the Compton scales.  A common route to reduce
relativistic kinetics uses a Foldy--Wouthuysen (FW) transformation (see e.g. \cite{silenko2015general,hinschberger2016coherent,ekman2017relativistic}) to obtain
a Hamiltonian for a selected energy sector, corresponding to electrons or positrons. The resulting effective
Hamiltonian keeps quantum corrections within that sector. A different
objective, which is adopted in the present paper, is to reduce the DHW equations while keeping the quantum
terms at a prescribed order, including terms that couple the
approximately identified electron and positron distributions. This
distinction is particularly relevant when the vacuum response is to be
kept together with the dynamics of real particles.

An electrostatic realization of this approach, limited to a 1D geometry, was recently developed in Ref. ~\cite{lundstrom2026modified}. Weak-field
and slow-time expansions resulted in electron and positron distribution
functions coupled through quantum corrections. The resulting description
contained both vacuum polarization and additional contributions that
depend on the real-particle distributions. This provides a motivation
for extending the reduction scheme to general electromagnetic fields and a full 3D geometry, where
spin and magnetization variables must also be included.

In the present work, we derive such a reduced model for the general 3D case as described above. By successively eliminating the DHW phase-space functions, we reduce the original 16 variables to eight independent scalar quantities. A further change of dependent variables expresses the remaining functions as one scalar and one three-component vector for each particle species of the Dirac field. We refer to this description as the weakly coupled electron--positron (WCEP) limit of the DHW equations. The transport operators separate into two semi-classical relativistic operators, while the quantum source terms keep the coupling between them.
Further approximations reduce the system to a description of a single species. In that
limit, we recover the well-known magnetic dipole terms and show how
the vector variables are related to rest-frame BMT precession, including
the Thomas factor (see e.g. Ref. \cite{jackson1999classical}). Furthermore, we demonstrate that the elimination procedure can be continued to all orders. This allows vacuum contributions to be kept in the reduced description and in
the reconstructed electromagnetic sources. We discuss how the procedure can be used to recover the effect
of vacuum polarization, including the need for
a charge renormalization.  

The paper is organized as follows. Sec.~\ref{sec:DHW} summarizes the
DHW formalism and the operators used in the expansion.
Sec.~\ref{sec:derivations} derives the reduced scalar and vector
equations and compares their single-species limits with established
spin kinetic models. Sec.~\ref{sec:all_orders_vacuum} describes the
extension to arbitrary perturbative order and the inclusion of vacuum
polarization. The final section summarizes the results and discusses
their implications.

\section{The Dirac--Heisenberg--Wigner formalism}
\label{sec:DHW}

In this section we give an introduction to the Dirac--Heisenberg--Wigner
(DHW) formalism, see e.g. Refs.  \cite{BialynickiBirula1991,hebenstreit2010schwinger,brodin2022quantum}. The theory is based on the Dirac equation,
and use a Wigner transform to obtain a phase-space description. Unlike theories based on the
Schr\"odinger or Pauli equations, the electron and positron degrees of
freedom are not separated from the start. Thus the same set of
variables can describe plasma dynamics as well as vacuum polarization
and electron--positron pair creation. For a discussion of radiation reaction, which lies outside the present DHW model, see Ref.~\cite{gonoskov2022charged}.  

The starting point is a gauge-invariant Wigner transformation of the
equal-time density operator. Since the Dirac wave function has four
components, the Wigner function is a $4\times4$ matrix. Expanding this
matrix in the generators of the Clifford algebra gives 16 real
phase-space functions. Following Ref.~\cite{BialynickiBirula1991}, we
write these as four scalars, $f_0,f_1,f_2,f_3$ and four three-vectors, $\mathbf g_0,\mathbf g_1,\mathbf g_2,\mathbf g_3$,
where all functions depend on $(\mathbf r,\mathbf p,t)$, with
$\mathbf r$, $\mathbf p$, and $t$ denoting position, kinetic momentum,
and time, respectively. Furthermore, we will use $p=|\mathbf p|$.
The phase-space variables
have clear physical interpretations, although they cannot be viewed as ordinary particle
distribution functions. For example, $f_0$ is associated with charge
density, $\mathbf{g}_1$ with current density, $f_3$ with mass density, $\mathbf g_0$ with spin density, and
$\mathbf g_3$ with magnetization density, see e.g. Ref. \cite{BialynickiBirula1991} for further elaborations. We stress that none of the
individual functions describes electrons or positrons separately.

We will use SI units throughout. Moreover, we denote the electric and magnetic fields by $\mathbf E(\mathbf r,t)$
and $\mathbf B(\mathbf r,t)$, $m$ is the electron
rest mass, $c$ is the speed of light, and $\hbar$ is the reduced Planck
constant. The gradients $\nabla_r$ and $\nabla_p$ act with respect to
position and momentum, respectively, and $\partial_t=\partial/\partial t$.

Treating the electromagnetic field as a classical mean field, i.e.,
using the Hartree approximation, the DHW equations are written as
\begin{subequations}
\label{eq:DHW}
\begin{align}
 D_t f_0+c\mathbf D\cdot\mathbf g_1 &=0,
 \label{eq:DHW_f0}\\
 D_t f_1+c\mathbf D\cdot\mathbf g_0
    &=-\frac{2mc^2}{\hbar}f_2,
 \label{eq:DHW_f1}\\
 D_t f_2+\frac{2c}{\hbar}\mathbf P\cdot\mathbf g_3
    &=\frac{2mc^2}{\hbar}f_1,
 \label{eq:DHW_f2}\\
 D_t f_3-\frac{2c}{\hbar}\mathbf P\cdot\mathbf g_2 &=0,
 \label{eq:DHW_f3}\\
 D_t\mathbf g_0+c\mathbf D f_1
    -\frac{2c}{\hbar}\mathbf P\times\mathbf g_1 &=0,
 \label{eq:DHW_g0}\\
 D_t\mathbf g_1+c\mathbf D f_0
    -\frac{2c}{\hbar}\mathbf P\times\mathbf g_0
    &=-\frac{2mc^2}{\hbar}\mathbf g_2,
 \label{eq:DHW_g1}\\
 D_t\mathbf g_2+c\mathbf D\times\mathbf g_3
    +\frac{2c}{\hbar}\mathbf P f_3
    &=\frac{2mc^2}{\hbar}\mathbf g_1,
 \label{eq:DHW_g2}\\
 D_t\mathbf g_3-c\mathbf D\times\mathbf g_2
    -\frac{2c}{\hbar}\mathbf P f_2 &=0.
 \label{eq:DHW_g3}
\end{align}
\end{subequations}
Here $e$ is the charge used in the Dirac equation \footnote{Note that the DHW-equations will hold as is for both possible sign conventions, that is, independently of whether we pick $e$ as the negative electron charge or as the positive elementary charge}. The nonlocal operators appearing above are given by
\begin{align}
D_t &=
 \frac{\partial}{\partial t}
 +e\int_{-1/2}^{1/2}d\lambda\,
 \mathbf E\left(
      \mathbf r+i\hbar\lambda\nabla_p,t
         \right)\cdot\nabla_p ,
\label{eq:Dt_full}\\
\mathbf D &=
 \nabla_r
 +e\int_{-1/2}^{1/2}d\lambda\,
 \mathbf B\left(
     \mathbf r+i\hbar\lambda\nabla_p,t
         \right)\times\nabla_p ,
\label{eq:D_full}\\
\mathbf P &=
 \mathbf p
 -ie\hbar\int_{-1/2}^{1/2}d\lambda\,\lambda\,
 \mathbf B\left(
     \mathbf r+i\hbar\lambda\nabla_p,t
         \right)\times\nabla_p .
\label{eq:P_full}
\end{align}

The shifted field arguments are a short way of writing a Taylor series.
Accordingly, expanding the electric field gives spatial derivatives of
$\mathbf E$, multiplied by momentum derivatives acting on the DHW functions.

For the derivations in the sections to follow, we separate the local contributions from the
first nonlocal corrections. Expanding the operators gives
\begin{align}
D_t &=
\underbrace{\left(
 \frac{\partial}{\partial t}
 +e\mathbf E\cdot\nabla_p
          \right)}_{D_{t,\mathrm{loc}}}
-\frac{e\hbar^2}{24}
 \left(\nabla_r\cdot\nabla_p\right)^2
 \mathbf E\cdot\nabla_p
+\mathcal O(\hbar^4),
\label{eq:Dt_expansion}\\
\mathbf D &=
\underbrace{\left(
 \nabla_r+e\mathbf B\times\nabla_p
          \right)}_{\mathbf D_{\mathrm{loc}}}
-\frac{e\hbar^2}{24}
 \left(\nabla_r\cdot\nabla_p\right)^2
 \mathbf B\times\nabla_p
+\mathcal O(\hbar^4),
\label{eq:D_expansion}\\
\mathbf P &=
\underbrace{\mathbf p}_{\mathbf P_{\mathrm{loc}}}
+\frac{e\hbar^2}{12}
 \left(\nabla_r\cdot\nabla_p\right)
 \mathbf B\times\nabla_p
+\mathcal O(\hbar^4).
\label{eq:P_expansion}
\end{align}
In these expressions, $\nabla_r$ acts on the field immediately following
it, whereas $\nabla_p$ acts on the phase-space function to the right of
the operator. For scale lengths much longer than the relevant de Broglie
wavelength, we can use the local approximation,
\begin{equation}
 D_t\rightarrow D_{t,\mathrm{loc}},\qquad
 \mathbf D\rightarrow\mathbf D_{\mathrm{loc}},\qquad
 \mathbf P\rightarrow\mathbf p.
\label{eq:local_operators}
\end{equation}
The corrections to this approximation describe quantum effects associated with the dispersive properties of the wave-function. Below we will keep these corrections to a given order. 

Together with Maxwell's equations, the DHW equations constitute a closed system. In the DHW formalism, the source terms in Maxwell's equations due to the Dirac field are  given by 
\begin{equation}
 \rho=\frac{e}{(2\pi\hbar)^3}\int f_0 d^3p
\label{eq:DHW_charge_density}
\end{equation}
and 
\begin{equation}
 \mathbf{j}=\frac{ec}{(2\pi\hbar)^3}\int \mathbf{g}_1 d^3p
\label{eq:DHW_current_density}
\end{equation}
respectively, where $\rho$ is the charge density, and $\mathbf{j}$ is the current density. The factor $(2\pi\hbar)^{-3}$ is the
phase-space normalization for the dimensionless DHW functions used here. 

A key difference from classical kinetic theory is that the DHW functions do not need to vanish when no real particles are present. They are
expectation values of field operators, and have nonzero vacuum values.
For the unperturbed vacuum these are given by
\begin{align}
 f_{3,\mathrm{vac}}(\mathbf p)
   &=-\frac{2mc^2}{\epsilon_p},
 \label{eq:f3vac}\\
 \mathbf g_{1,\mathrm{vac}}(\mathbf p)
   &=-\frac{2c\mathbf p}{\epsilon_p},
 \label{eq:g1vac}
\end{align}
where
\begin{equation}
 \epsilon_p=\sqrt{m^2c^4+c^2p^2},
\label{eq:energy}
\end{equation}
is the particle energy. All other components vanish for a vacuum that is not spin-polarized \footnote{Note, however, that a strong magnetic field may lead to significant spin-polarization of the vacuum, in which case additional DHW-functions will have nonzero vacuum values}.  The expressions
above follow from evaluating the Wigner operator in the free-field vacuum.
Note that the momentum integral of $\mathbf g_{1,\mathrm{vac}}$ vanishes as the phase-space vacuum current density is an odd function of momentum. 

In general, the phase-space functions contain both particle and vacuum contributions. When we keep the vacuum response, however,
ultraviolet divergences typically occur, and renormalization is needed
\cite{BialynickiBirula1991,fedotov2023advances,lundstrom2026modified}, as we will come back to in Sec.~\ref{subsec:ao_vacuum_response}. 

In the present work, we will use the fact that the DHW equations contain both positive- and negative-energy degrees of freedom, corresponding to positrons and electrons. In the weak-quantum
limit, suitable combinations of $f_i$ and $\mathbf g_i$ can approximately
separate the two species. As we will see, somewhat different combinations are
needed for the scalar and vector functions. Moreover, the spin variables
obtained in this way differ from the usual rest-frame spin distribution.

\section{Derivations of the WCEP equations}
In this section we present a systematic expansion of the DHW equations, aiming to find an approximate separation of the electron and positron degrees of freedom. In  Sec.~\ref{sec:scalar_evolution}, we perform the separation for the scalar phase-space variables, and in Sec.~\ref{subsec:spin_evolution} we do the same for the vector variables of the DHW formalism. Finally, in Sec.~\ref{sec:Comparison} we demonstrate agreement with previous works in appropriate limiting cases.    

\label{sec:derivations}
\subsection{Scalar evolution equations}
\label{sec:scalar_evolution}

We first derive the scalar equations, keeping quantum corrections up to
second order. The momentum dependence is kept fully relativistic, i.e.,
we do not expand in $p/(mc)$. In addition to the expansion of 
operators, we assume weak spin polarization, i.e., first order in a weak quantum expansion. Weak spin polarization is not required by the general procedure; we assume it here to determine which quantum coupling terms to keep at each order. The above scaling gives us the ordering
\begin{equation}
\begin{aligned}
 \mathbf g_0,\ \mathbf g_3&=\mathcal O(\hbar),\\
 f_0,\ f_3,\ \mathbf g_1&=\mathcal O(1).
\end{aligned}
 \label{eq:scalar_ordering}
\end{equation}
The powers of $\hbar$ here specify the order of the expansion. As will become apparent below, independent oscillations at the Compton
time scale $\hbar/(mc^2)$ are eliminated by the expansion procedure. Next, we write the operators as
\begin{equation}
\begin{aligned}
 D_t&=D_{t,\mathrm{loc}}+\delta D_t,\\
 \mathbf D&=\mathbf D_{\mathrm{loc}}+\delta\mathbf D,\\
 \mathbf P&=\mathbf p+\delta\mathbf P,
\end{aligned}
 \label{eq:scalar_operator_split}
\end{equation}
where the corrections are the terms in
Eqs.~\eqref{eq:Dt_expansion}--\eqref{eq:P_expansion} of order $\hbar^2$. Operator products
act successively on the expression to their right. Note that momentum
derivatives act on explicit momentum-dependent factors as well as on the
phase-space functions.

From Eqs.~\eqref{eq:DHW_g1} and \eqref{eq:DHW_g2} we immediately obtain
\begin{align}
 \mathbf g_2 &=
 \frac{1}{mc}\mathbf P\times\mathbf g_0\nonumber\\
 &\quad -\frac{\hbar}{2mc^2}
     \left(D_t\mathbf g_1+c\mathbf D f_0\right),
 \label{eq:scalar_g2_exact}\\
 \mathbf g_1 &=
 \frac{1}{mc}\mathbf P f_3\nonumber\\
 &\quad +\frac{\hbar}{2mc^2}
     \left(D_t\mathbf g_2+c\mathbf D\times\mathbf g_3\right).
 \label{eq:scalar_g1_exact}
\end{align}
We solve these relations for $\mathbf{g}_1$ and $\mathbf{g}_2$ by successive substitutions. Inserting the
lowest-order result $\mathbf g_1=\mathbf p f_3/(mc)$ gives the
first-order expression for $\mathbf g_2$, denoted by $\mathbf g_2^{(1)}$,
where the superscript indicates truncation at first order:
\begin{equation}
\begin{aligned}
 \mathbf g_2^{(1)}={}&\frac{\mathbf p\times\mathbf g_0}{mc}\\
 &-\frac{\hbar}{2mc^2}\left[
 D_{t,\mathrm{loc}}\left(\frac{\mathbf p f_3}{mc}\right)
 +c\mathbf D_{\mathrm{loc}}f_0\right].
\end{aligned}
 \label{eq:scalar_K}
\end{equation}
Here the time derivative also acts on $\mathbf p$, giving
$D_{t,\mathrm{loc}}\mathbf p=e\mathbf E$. Inserting this expression
for $\mathbf g_2$ back into Eq.~\eqref{eq:scalar_g1_exact}, we find, to second order,
\begin{equation}
\begin{aligned}
 \mathbf g_1&=\frac{\mathbf p f_3}{mc}+\delta\mathbf g_1,\\
 \delta\mathbf g_1&=\frac{\delta\mathbf P f_3}{mc}\\
 &\quad+\frac{\hbar}{2mc^2}\left(
 D_{t,\mathrm{loc}}\mathbf g_2^{(1)}
 +c\mathbf D_{\mathrm{loc}}\times\mathbf g_3\right).
\end{aligned}
 \label{eq:scalar_J}
\end{equation}
Thus $\mathbf g_2^{(1)}$ is of order $\hbar$, and the correction
$\delta\mathbf g_1$ is of order $\hbar^2$. 

Substitution into Eq.~\eqref{eq:DHW_f0}, using  $\mathbf D_{\mathrm{loc}}\cdot(\mathbf p f_3)
=\mathbf p\cdot\mathbf D_{\mathrm{loc}}f_3$, now gives
\begin{equation}
 D_{t,\mathrm{loc}}f_0+
 \frac{\mathbf p}{m}\cdot\mathbf D_{\mathrm{loc}}f_3
 =Q_0,
 \label{eq:scalar_f0_reduced}
\end{equation}
with
\begin{equation}
\begin{aligned}
 Q_0={}&-\delta D_t f_0
 -\frac{1}{m}\delta\mathbf D\cdot(\mathbf p f_3)\\
 &-c\mathbf D_{\mathrm{loc}}\cdot\delta\mathbf g_1.
\end{aligned}
 \label{eq:scalar_Q0}
\end{equation}

For the second scalar equation, we insert
Eq.~\eqref{eq:scalar_g2_exact} into Eq.~\eqref{eq:DHW_f3}.
Before expanding the operators, the result is
\begin{equation}
\begin{aligned}
 D_t f_3={}&\frac{2}{m\hbar}\mathbf P\cdot
 (\mathbf P\times\mathbf g_0)\\
 &-\frac{1}{mc}\mathbf P\cdot D_t\mathbf g_1
 -\frac{1}{m}\mathbf P\cdot\mathbf D f_0.
\end{aligned}
 \label{eq:scalar_f3_intermediate}
\end{equation}
The first term is zero if $\mathbf P$ is replaced by $\mathbf p$.
However, the nonlocal correction gives
\begin{equation}
\begin{aligned}
 \mathbf P\cdot(\mathbf P\times\mathbf g_0)
 ={}&\mathbf p\cdot(\delta\mathbf P\times\mathbf g_0)\\
 &+\delta\mathbf P\cdot(\mathbf p\times\mathbf g_0)\\
 &+\text{higher-order terms}.
\end{aligned}
 \label{eq:scalar_P_cross}
\end{equation}
Since $\mathbf g_0$ is of order $\hbar$, the factor $1/\hbar$ in
Eq.~\eqref{eq:scalar_f3_intermediate} makes this contribution second
order. It must therefore be kept. 
Inserting Eq.~\eqref{eq:scalar_J} and moving the local terms to the
left-hand side gives
\begin{equation}
\begin{aligned}
 \gamma_p^2D_{t,\mathrm{loc}}f_3
 &+\frac{e\mathbf E\cdot\mathbf p}{m^2c^2}f_3\\
 &+\frac{\mathbf p}{m}\cdot\mathbf D_{\mathrm{loc}}f_0=Q_3,\\
 \gamma_p&=\frac{\epsilon_p}{mc^2}
 =\sqrt{1+\frac{p^2}{m^2c^2}},
\end{aligned}
 \label{eq:scalar_f3_reduced}
\end{equation}
where the quantum corrections on the right-hand side are
\begin{equation}
\begin{aligned}
 Q_3={}&-\delta D_t f_3\\
 &-\frac{1}{mc}\delta\mathbf P\cdot
 D_{t,\mathrm{loc}}\left(\frac{\mathbf p f_3}{mc}\right)\\
 &-\frac{1}{mc}\mathbf p\cdot\delta D_t
 \left(\frac{\mathbf p f_3}{mc}\right)\\
 &-\frac{1}{mc}\mathbf p\cdot D_{t,\mathrm{loc}}\delta\mathbf g_1\\
 &-\frac{1}{m}\delta\mathbf P\cdot\mathbf D_{\mathrm{loc}}f_0
 -\frac{1}{m}\mathbf p\cdot\delta\mathbf D f_0\\
 &+\frac{2}{m\hbar}\mathbf p\cdot
 (\delta\mathbf P\times\mathbf g_0)\\
 &+\frac{2}{m\hbar}\delta\mathbf P\cdot
 (\mathbf p\times\mathbf g_0).
\end{aligned}
\label{eq:scalar_Q3}
\end{equation}
Both $Q_0$ and $Q_3$ are of second order. Time derivatives of the scalar
functions inside these corrections can therefore be replaced using
Eqs.~\eqref{eq:scalar_f0_reduced} and \eqref{eq:scalar_f3_reduced}
with $Q_0=Q_3=0$. Naturally, this replacement does not change the accuracy of the calculation scheme and avoids introducing extra evolution equations.

To make contact with classical kinetic theory, we introduce the variables
\begin{equation}
 f_\pm=\frac{1}{2}\left(\gamma_p f_3\pm f_0\right) \label{eq:scalar_fpm} 
 \end{equation}
 with the inverse relations
 \begin{equation}
    f_0=f_+-f_-,\qquad
 f_3=\frac{f_++f_-}{\gamma_p}.
  \label{eq:scalar_fpm_inverse}
\end{equation} 
 This change of variables makes the leading-order operators (the parts surviving in the classical limit) of Eqs.~\eqref{eq:scalar_f0_reduced} and \eqref{eq:scalar_f3_reduced} identical. Moreover,
 with $\mathbf v_p=\mathbf p/(m\gamma_p)=c^2\mathbf p/\epsilon_p$,
the two operators are
\begin{equation}
\begin{aligned}
 D_{v,\pm}
 &=D_{t,\mathrm{loc}}\pm\mathbf v_p\cdot\mathbf D_{\mathrm{loc}}\\
 &=\frac{\partial}{\partial t}
 \pm\mathbf v_p\cdot\nabla_r\\
 &\quad+e\left(\mathbf E\pm\mathbf v_p\times\mathbf B\right)
 \cdot\nabla_p,
\end{aligned}
 \label{eq:scalar_Vlasov}
\end{equation}
coinciding with the classical relativistic Vlasov operator. To fully appreciate this result, one must recognize the built-in feature of the DHW formalism: for electrons and positrons moving in the same direction, the momentum variable $\mathbf p$ has opposite signs; see, e.g., Ref. \cite{brodin2022quantum}. Thus $e\left(\mathbf E\pm\mathbf v_p\times\mathbf B\right)$ represents the classical Lorentz force expression, independent of the sign in front of the magnetic force term.   

Next, when combining the scalar equations, we use
\begin{equation}
\begin{aligned}
 D_{t,\mathrm{loc}}\gamma_p
 &=\frac{e\mathbf E\cdot\mathbf p}{m^2c^2\gamma_p},\\
 \mathbf p\cdot\mathbf D_{\mathrm{loc}}\gamma_p&=0.
\end{aligned}
 \label{eq:scalar_gamma_identities}
\end{equation}
Dividing Eq.~\eqref{eq:scalar_f3_reduced} by $\gamma_p$, adding
or subtracting Eq.~\eqref{eq:scalar_f0_reduced}, and dividing by two,
we obtain
\begin{equation}
 D_{v,\pm}f_\pm
 =\frac{1}{2}\left(\frac{Q_3}{\gamma_p}\pm Q_0\right).
 \label{eq:scalar_fpm_evolution}
\end{equation}
The right-hand side is specified by
Eqs.~\eqref{eq:scalar_K}, \eqref{eq:scalar_J}, \eqref{eq:scalar_Q0},
and \eqref{eq:scalar_Q3}, with $f_0$ and $f_3$ expressed through
$f_\pm$ using Eq.~\eqref{eq:scalar_fpm_inverse}. In the classical limit $Q_0=Q_3=0$, and the equations
reduce to $D_{v,\pm}f_\pm=0$, as expected.

In Eq. (\ref{eq:scalar_fpm_evolution}), we have chosen to keep terms to second order in $\hbar$. The expansion in spatial nonlocality is distinct from the slow-time expansion used to eliminate the auxiliary DHW functions. In principle, the latter procedure can be performed while keeping the full spatially nonlocal operators, without assuming their nonlocal parts to be small. The scalar equations can still be written with the left-hand side of Eq.~\eqref{eq:scalar_fpm_evolution}, but their right-hand sides must then include contributions omitted in the present spatial-gradient truncation, including products of nonlocal corrections. The general elimination procedure is discussed in Sec.~\ref{sec:all_orders_vacuum}. Moreover, if we take the opposite limit (where quantum corrections proportional to the de Broglie length over characteristic scale lengths are negligible compared to quantum corrections associated with the spin), the coupling term in Eq. (\ref{eq:scalar_fpm_evolution}) reduces to
\begin{equation}
\begin{aligned}
 &\frac12\left(\frac{Q_3}{\gamma_p}\pm Q_0\right)
 \\
 &=-\frac{\hbar}{4m}\Biggl[
 \frac{1}{c^2}
 \left(
 \frac{\mathbf p}{mc\gamma_p}\cdot D_{t,\mathrm{loc}}
 \pm c\mathbf D_{\mathrm{loc}}\cdot
 \right)
 \\
 &\quad\times\Biggl\{
 D_{t,\mathrm{loc}}
 \left(\frac{\mathbf p\times\mathbf g_0}{mc}\right)
 +c\mathbf D_{\mathrm{loc}}\times\mathbf g_3
 \\
 &\quad-\frac{\hbar}{2mc^2}D_{t,\mathrm{loc}}
 \Biggl[
 D_{t,\mathrm{loc}}
 \left(\frac{\mathbf p f_3}{mc}\right)
 \\
 &\quad+c\mathbf D_{\mathrm{loc}}f_0
 \Biggr]\Biggr\}
 +\frac{e}{\gamma_p}
 (\partial_{r_j}B_i)\partial_{p_j}g_{0i}
 \Biggr].
\end{aligned}
\label{eq:scalar_fpm_local_source}
\end{equation}
The above expression will be referred to as the semi-classical limit. This approximation keeps the spin-dependent contributions at the order considered, including the magnetic-gradient coupling, while omitting the orbital dispersive corrections. As we will see in Sec.~\ref{sec:Comparison} when comparing with previous works, in the weakly relativistic limit Eq. (\ref{eq:scalar_fpm_local_source}) reduces considerably, and that result coincides with previous works. 

\subsection{Spin and magnetization evolution equations}
\label{subsec:spin_evolution}

Next we derive the vector equations, using the same slow-time assumption
and ordering as above. Since $\mathbf g_0$ and $\mathbf g_3$ are of
order $\hbar$, we keep their leading evolution, including scalar source
terms of that order. Again, no expansion in $p/(mc)$ is made. We use
$D_{t,\mathrm{loc}}$ and $\mathbf D_{\mathrm{loc}}$ as defined in
Sec.~\ref{sec:DHW}. The nonlocal correction to $\mathbf P$ must still be kept in terms
containing a factor $1/\hbar$.

Using Eqs. (\ref{eq:DHW_f1})-(\ref{eq:DHW_f2}), we write
\begin{align}
 f_1&=\frac{\mathbf P\cdot\mathbf g_3}{mc}
     +\frac{\hbar}{2mc^2}D_t f_2,
 \label{eq:spin_f1_exact}\\
 f_2&=-\frac{\hbar}{2mc^2}
     \left(D_t f_1+c\mathbf D\cdot\mathbf g_0\right).
 \label{eq:spin_f2_exact}
\end{align}
It follows that $f_1=\mathbf p\cdot\mathbf g_3/(mc)
+\mathcal O(\hbar^3)$ and $f_2=\mathcal O(\hbar^2)$, although division by $\hbar$ in Eq. (\ref{eq:DHW_g3}) means we must retain $f_2$ to the order $\hbar^2$ here.  
Together with the substitution for $\mathbf g_1$ and $\mathbf g_2$
used above, this gives
\begin{align}
 \mathbf g_2={}&\frac{\mathbf p\times\mathbf g_0}{mc}
     -\frac{\hbar}{2mc^2}\mathbf U
     +\mathcal O(\hbar^3),
 \label{eq:spin_g2}\\
 \mathbf g_1={}&\frac{(\mathbf p+\delta\mathbf P)f_3}{mc}
     +\frac{\hbar}{2mc^2}
     \left[\frac{D_{t,\mathrm{loc}}(\mathbf p\times\mathbf g_0)}{mc}
          +c\mathbf D_{\mathrm{loc}}\times\mathbf g_3\right]\nonumber\\
 &-\frac{\hbar^2}{4m^2c^4}D_{t,\mathrm{loc}}\mathbf U
     +\mathcal O(\hbar^4),
 \label{eq:spin_g1}\\
 \mathbf U={}&D_{t,\mathrm{loc}}\left(\frac{\mathbf p f_3}{mc}\right)
              +c\mathbf D_{\mathrm{loc}} f_0.
 \label{eq:spin_U_initial}
\end{align}
Here $\delta\mathbf P$ is the second-order correction in
Eq.~\eqref{eq:P_expansion}. Note that the last displayed term in
$\mathbf g_1$ cannot be dropped: although it is of second order, the factor $1/\hbar$ in the 
$\mathbf g_0$ equation, Eq. (\ref{eq:DHW_g0}), makes its contribution
first order. 

Inserting these results into the remaining vector equations gives
\begin{align}
 D_{t,\mathrm{loc}}\mathbf g_0={}&
 \frac{\mathbf p\times D_{t,\mathrm{loc}}(\mathbf p\times\mathbf g_0)}{m^2c^2}
 +\frac{\mathbf p\times(\mathbf D_{\mathrm{loc}}\times\mathbf g_3)}{m}
 \nonumber\\
 &-\frac{\mathbf D_{\mathrm{loc}}(\mathbf p\cdot\mathbf g_3)}{m}
 +\mathbf S_0,
 \label{eq:spin_g0_intermediate}\\
 D_{t,\mathrm{loc}}\mathbf g_3={}&\frac{\mathbf D_{\mathrm{loc}}\times(\mathbf p\times\mathbf g_0)}{m}
 -\frac{\mathbf p}{m}\mathbf D_{\mathrm{loc}}\cdot\mathbf g_0
 \nonumber\\
 &-\frac{\mathbf p}{m^2c^2}D_{t,\mathrm{loc}}(\mathbf p\cdot\mathbf g_3)
 +\mathbf S_3,
 \label{eq:spin_g3_intermediate}
\end{align}
where we have introduced the source terms
\begin{align}
 \mathbf S_0={}&\frac{2}{m\hbar}
      \mathbf P\times(\mathbf P f_3)
      -\frac{\hbar}{2m^2c^3}\mathbf p\times D_{t,\mathrm{loc}}\mathbf U,
 \label{eq:spin_S0_initial}\\
 \mathbf S_3={}&-\frac{\hbar}{2mc}\mathbf D_{\mathrm{loc}}\times\mathbf U.
 \label{eq:spin_S3_initial}
\end{align}
Only the first nonlocal contribution to the $\mathbf P\times
(\mathbf P f_3)$ term is needed. Writing
$\partial_{r_j}=\partial/\partial r_j$ and
$\partial_{p_j}=\partial/\partial p_j$, we obtain
\begin{equation}
\begin{aligned}
 \mathbf P\times(\mathbf P f_3)
 ={}&\mathbf p\times(\delta\mathbf P f_3)
     +\delta\mathbf P\times(\mathbf p f_3)\\
 ={}&\frac{e\hbar^2}{12}
     \bigl[(\nabla_r\cdot\mathbf B)\nabla_p f_3\\
 &\qquad-3(\partial_{r_j}\mathbf B)\partial_{p_j}f_3\bigr]
     +\mathcal O(\hbar^4),
\end{aligned}
\label{eq:spin_P_cross_P}
\end{equation}
where Cartesian indices run over the three spatial components and
repeated Cartesian indices are summed over. Using 
$\nabla_r\cdot\mathbf B=0$, the source becomes
\begin{equation}
 \frac{2}{m\hbar}\mathbf P\times(\mathbf P f_3)
 =-\frac{e\hbar}{2m}
     (\partial_{r_j}\mathbf B)\partial_{p_j}f_3
     +\mathcal O(\hbar^3).
\label{eq:spin_magnetic_gradient_source}
\end{equation}
To deal with terms that combine momentum-dependent factors with time derivatives, it will be useful to introduce the following notation
\begin{equation}
\begin{aligned}
 N_{ij}&=\delta_{ij}+\frac{p_i p_j}{m^2c^2},\\
 M_{ij}&=\gamma_p^2\delta_{ij}-\frac{p_i p_j}{m^2c^2},\\
 M_{ij}N_{jk}&=\gamma_p^2\delta_{ik},
\end{aligned}
\label{eq:spin_matrices}
\end{equation}
where $\delta_{ij}$ is the Kronecker delta. We use $\epsilon_{ijk}$ for the Levi-Civita symbol. The relations $D_{t,\mathrm{loc}}p_i=eE_i$ and
$D_{\mathrm{loc},i}p_j=e\epsilon_{ikj}B_k$ give
\begin{equation}
\begin{aligned}
 \mathbf p\times(\mathbf D_{\mathrm{loc}}\times\mathbf g)
   -\mathbf D_{\mathrm{loc}}(\mathbf p\cdot\mathbf g)
   & =-(\mathbf p\cdot\mathbf D_{\mathrm{loc}})\mathbf g
      -e\mathbf B\times\mathbf g,\\
 \mathbf D_{\mathrm{loc}}\times(\mathbf p\times\mathbf g)
   -\mathbf p(\mathbf D_{\mathrm{loc}}\cdot\mathbf g)
   & =-(\mathbf p\cdot\mathbf D_{\mathrm{loc}})\mathbf g
      -e\mathbf B\times\mathbf g.
\end{aligned}
\label{eq:spin_product_identities}
\end{equation}
Here $\mathbf g$ denotes an arbitrary vector phase-space function.

Collecting the terms, we can now write the coupled equations in
component form as
\begin{align}
 M_{ij}D_{t,\mathrm{loc}}g_{0j}
 &+\frac{p_j}{m}D_{\mathrm{loc},j}g_{3i}
 +\frac{e}{m}(\mathbf B\times\mathbf g_3)_i\nonumber\\
 &-\frac{e}{m^2c^2}[\mathbf p\times(\mathbf E\times\mathbf g_0)]_i
 =S_{0i},\label{eq:spin_g0_reduced}\\
 N_{ij}D_{t,\mathrm{loc}}g_{3j}
 &+\frac{p_j}{m}D_{\mathrm{loc},j}g_{0i}
 +\frac{e}{m}(\mathbf B\times\mathbf g_0)_i\nonumber\\
 &+\frac{ep_i}{m^2c^2}E_jg_{3j}
 =S_{3i}.\label{eq:spin_g3_reduced}
\end{align}

The spatial transport, the magnetic force term, and spin precession all couple
$\mathbf g_0$ to $\mathbf g_3$. To separate the two transport equations into electron and positron degrees of freedom, we introduce variables that are momentum-dependent linear combinations of the spin and magnetization vectors, according to
\begin{equation}
 F_{\pm i}=g_{0i}\pm A_{ij}g_{3j}
 \label{eq:spin_Fpm_definition}
 \end{equation}
 where
 \begin{equation}
   A_{ij}=\frac{1}{\gamma_p}\left(\delta_{ij}
                   +\frac{p_i p_j}{m^2c^2}\right)   
 \label{eq:spin_A_definition}
\end{equation}
 with the inverse
 \begin{equation}
 A^{-1}_{ij}=\gamma_p\delta_{ij}
                     -\frac{p_i p_j}{m^2c^2\gamma_p}.
 \label{eq:spin_A_inverse}
\end{equation}
Inverting the change of variables gives us
\begin{equation}
\begin{aligned}
 g_{0i}&=\frac{F_{+i}+F_{-i}}{2},\\
 g_{3i}&=\frac12\left(\gamma_p\delta_{ij}
            -\frac{p_i p_j}{m^2c^2\gamma_p}\right)(F_{+j}-F_{-j}).
\end{aligned}
\label{eq:spin_Fpm_inverse}
\end{equation}
Since the coefficients $A_{ij}$ in Eq.~\eqref{eq:spin_A_definition} depend on momentum, their derivatives must
be included when completing this change of variables. In particular, we note that
\begin{equation}
\begin{aligned}
 D_{t,\mathrm{loc}}A_{ij}
 ={}&\frac{e(E_i p_j+p_i E_j)}{m^2c^2\gamma_p}\\
 &-\frac{eE_kp_k}{m^2c^2\gamma_p^3}N_{ij},\\
 p_kD_{\mathrm{loc},k}A_{ij}
 ={}&-e\epsilon_{i\ell k}B_\ell A_{kj}
       +e A_{ik}\epsilon_{k\ell j}B_\ell.
\end{aligned}
\label{eq:spin_A_derivatives}
\end{equation}
Next, we multiply Eq. (\ref{eq:spin_g0_reduced}) by $N_{ij}/\gamma_p^2$, and Eq. (\ref{eq:spin_g3_reduced}) by $\pm\delta_{ij}/\gamma_p$, and add the results. Using the
relations above, the result is
\begin{equation}
\begin{aligned}
 D_{v,\pm}\mathbf F_\pm
 &\ \pm\frac{e}{m\gamma_p}\mathbf B\times\mathbf F_\pm\\
 &-\frac{e}{m^2c^2\gamma_p^2}
     \mathbf p\times(\mathbf E\times\mathbf F_\pm)
 =\boldsymbol{\mathcal S}_\pm,
\end{aligned}
\label{eq:spin_Fpm_evolution}
\end{equation}
where
\begin{equation}
 \mathcal S_{\pm i}
 =\frac{1}{\gamma_p^2}\left(\delta_{ij}
             +\frac{p_i p_j}{m^2c^2}\right)S_{0j}
       \pm\frac{S_{3i}}{\gamma_p},
\label{eq:spin_Fpm_source}
\end{equation}
and we have reused the Vlasov operator from Eq. (\ref{eq:scalar_Vlasov}). The key property of Eq. (\ref{eq:spin_Fpm_evolution}) is that the left-hand side contains a transport operator corresponding to semi-classical particle orbits for electrons/positrons, just as in Eq. (\ref{eq:scalar_fpm_evolution}), with the difference that the vector variables $\mathbf{F}_{\pm}$ are also subject to spin precession. More specifically,  
the second term of Eq. (\ref{eq:spin_Fpm_evolution}) describes magnetic spin precession. The term that follows is related to the Thomas precession contribution, but has a
double cross product, and does not in general preserve
$|\mathbf F_\pm|$. Thus $\mathbf F_\pm$ cannot be identified directly
with the usual rest-frame spin distribution, which is the reason the well-known expression for Thomas precession is hidden. We will return to this point
in Sec.~\ref{sec:Comparison} and show that Eq. (\ref{eq:spin_Fpm_evolution}) is nevertheless consistent with the well-known result regarding Thomas precession; see, e.g., Ref. \cite{jackson1999classical}.

We next give the scalar sources without time derivatives of the
phase-space functions. To do this, we use the leading scalar equations,
\begin{equation}
\begin{aligned}
 D_{t,\mathrm{loc}}f_0&=-\frac{p_j}{m}D_{\mathrm{loc},j}f_3,\\
 D_{t,\mathrm{loc}}f_3&=-\frac{1}{\gamma_p^2}
 \left(\frac{p_j}{m}D_{\mathrm{loc},j}f_0
             +\frac{eE_jp_j}{m^2c^2}f_3\right).
\end{aligned}
\label{eq:spin_UV}
\end{equation}

Substituting the second of these equations into the definition of
$\mathbf U$, we obtain
\begin{equation}
\begin{aligned}
 U_i&=c(N^{-1})_{ij}D_{\mathrm{loc},j}f_0+\frac{e}{mc}(N^{-1})_{ij}E_jf_3,\\
 (N^{-1})_{ij}&=\delta_{ij}-\frac{p_i p_j}{m^2c^2\gamma_p^2}.
\end{aligned}
\label{eq:spin_H_reduced}
\end{equation}
Next, we note that the commutator needed when differentiating $\mathbf U$ is
\begin{equation}
\begin{aligned}
 [D_{t,\mathrm{loc}},D_{\mathrm{loc},i}]
 ={}&e\epsilon_{ijk}(\partial_t B_j)\partial_{p_k}\\
 &-e(\partial_{r_i}E_j)\partial_{p_j}\\
 ={}&-e(\partial_{r_j}E_i)\partial_{p_j},
\end{aligned}
\label{eq:spin_Td_commutator}
\end{equation}
where the last equality follows from Faraday's law,
$\partial_t\mathbf B=-\nabla_r\times\mathbf E$.
The derivative of $(N^{-1})_{ij}$ is
\begin{equation}
\begin{aligned}
 D_{t,\mathrm{loc}}(N^{-1})_{ij}
 ={}&-\frac{e(E_i p_j+p_i E_j)}{m^2c^2\gamma_p^2}\\
 &+\frac{2eE_kp_k\,p_i p_j}{m^4c^4\gamma_p^4}.
\end{aligned}
\label{eq:spin_RE}
\end{equation}

Next, we use these expressions and replace the temporal derivatives of the scalars by Eq.~\eqref{eq:spin_UV}. We write
$U_{t,i}\equiv D_{t,\mathrm{loc}}U_i$ for the evaluated convective derivative, to distinguish it from $\partial_t U_i$.
Its components are
\begin{equation}
\begin{aligned}
 U_{t,i}={}&-c(N^{-1})_{ij}D_{\mathrm{loc},j}
           \left(\frac{p_k}{m}D_{\mathrm{loc},k}f_3\right)\\
 &-ec(N^{-1})_{ij}(\partial_{r_k}E_j)\partial_{p_k}f_0\\
 &+(D_{t,\mathrm{loc}}(N^{-1})_{ij})
           \left(cD_{\mathrm{loc},j}f_0+\frac{eE_j}{mc}f_3\right)\\
 &+\frac{e}{mc}(N^{-1})_{ij}(\partial_t E_j)f_3\\
 &-\frac{e(N^{-1})_{ij}E_j}{mc\gamma_p^2}
   \left(\frac{p_k}{m}D_{\mathrm{loc},k}f_0
             +\frac{eE_kp_k}{m^2c^2}f_3\right).
\end{aligned}
\label{eq:spin_W}
\end{equation}
This expression contains spatial and momentum derivatives of the scalar
functions, together with $\partial_t\mathbf E$. The replacement of
the scalar time derivatives changes the vector equations only at order
$\hbar^3$. The sources are consequently
\begin{align}
 \mathbf S_0={}&-\frac{e\hbar}{2m}
     (\partial_{r_j}\mathbf B)\partial_{p_j}f_3
     -\frac{\hbar}{2m^2c^3}\mathbf p\times\mathbf U_t,
 \label{eq:spin_S0_final}\\
 \mathbf S_3={}&-\frac{\hbar}{2mc}
     \bigl[\nabla_r\times\mathbf U
     +e\nabla_p(\mathbf B\cdot\mathbf U)\nonumber\\
 &\hspace{34mm}-e\mathbf B(\nabla_p\cdot\mathbf U)\bigr].
 \label{eq:spin_S3_final}
\end{align}
where we can use Eq. (\ref{eq:scalar_fpm_inverse}) to express $f_3$ and $f_0$ in terms of our new variables $f_{\pm}$.
Similarly, the electromagnetic sources in Maxwell's equations can be expressed in terms of
the new variables. Using the normalization of
Eqs.~\eqref{eq:DHW_charge_density} and
\eqref{eq:DHW_current_density}, the charge density is
\begin{equation}
 \rho=\frac{e}{(2\pi\hbar)^3}\int d^3p\,(f_+-f_-).
\label{eq:WCEP_charge}
\end{equation}
Furthermore, applying the same approximation scheme as when deriving the governing equations, the current source $\mathbf{g_1}$ can be expressed in terms of $f_{\pm}$ and $\mathbf{F}_{\pm}$. This can be done to any order in the expansion of $\mathbf{g}_1$, but below, for simplicity, we display only the transport current and the leading spin-dependent polarization and magnetization contributions. Spin-independent corrections of order $\hbar^2$ are omitted; these can be restored directly from the reconstructed expression for $\mathbf g_1$. With this additional approximation, the current density can be written as
\begin{equation}
 \mathbf j=\mathbf j_{\mathrm{tr}}
       +\partial_t\boldsymbol{\mathcal P}
       +\nabla_r\times\boldsymbol{\mathcal M},
\label{eq:WCEP_current}
\end{equation}
where the free current density is
\begin{equation}
 \mathbf j_{\mathrm{tr}}
 =\frac{e}{(2\pi\hbar)^3}\int d^3p\,\mathbf v_p(f_++f_-).
\label{eq:WCEP_transport_current}
\end{equation}
The polarization is
\begin{equation}
 \boldsymbol{\mathcal P}
 =\frac{e\hbar}{4m^2c^2(2\pi\hbar)^3}
       \int d^3p\mathbf p\times(\mathbf F_++\mathbf F_-)
\label{eq:WCEP_polarization}
\end{equation}
and the magnetization components are
\begin{equation}
 \mathcal M_i
 =\frac{e\hbar}{4m(2\pi\hbar)^3}
   \int d^3p\,A^{-1}_{ij}(F_{+j}-F_{-j}),
\label{eq:WCEP_magnetization}
\end{equation}
with $A^{-1}_{ij}$ given by Eq.~\eqref{eq:spin_A_inverse}.

Eqs. (\ref{eq:scalar_fpm_evolution}) and (\ref{eq:spin_Fpm_evolution}) are the main results of our paper. A consistently truncated closed system is obtained by combining these equations with Maxwell's equations and reconstructing the electromagnetic sources to the same order as the evolution equations. Naturally, we then also express the right-hand source terms in terms of the new variables $f_{\pm}$ and $\mathbf{F}_{\pm}$, using Eqs. (\ref{eq:scalar_fpm_inverse}) and (\ref{eq:spin_Fpm_inverse}). 

We will refer to this set as the weakly coupled electron--positron (WCEP) equations. The key feature is that the electron and positron degrees of freedom are separated, with the help of the variable changes made in Eqs. (\ref{eq:scalar_fpm}) and (\ref{eq:spin_Fpm_definition}). However, contrary to the case when a FW-transformation is made, there is a weak coupling between the electrons and positrons through the coupling terms in the right-hand sides of Eqs. (\ref{eq:scalar_fpm_evolution}) and (\ref{eq:spin_Fpm_evolution}). We note that one can approximate the system by keeping only one species (electrons or positrons) and disregarding the electron--positron cross-coupling, provided it is sufficiently weak to be dropped. However, we will show in Section IV that it is necessary to keep the electron--positron coupling, in order to recover well-known results involving vacuum polarization.  

\subsection{Comparison with previous works}
\label{sec:Comparison}

We first compare the non-relativistic spin terms with the semi-classical
model of Hurst \textit{et al.}~\cite{Hurst2014}. Next we compare the
relativistic precession with the usual rest-frame spin equation.
Throughout this subsection, we choose $e>0$ to be the elementary
charge and consider electrons, whose charge is $-e$. We omit the
vacuum contribution and the coupling to the positron sector, although
these contributions have been kept in the derivation above.

With this convention, electrons correspond to the minus branch.
We introduce their physical velocity through
\begin{equation}
\begin{aligned}
 \mathbf p&=-m\gamma\mathbf v,
 &\gamma&=\frac{1}{\sqrt{1-v^2/c^2}},\\
 f(\mathbf r,\mathbf v,t)
 &=f_-(\mathbf r,-m\gamma\mathbf v,t),\\
 \mathbf F(\mathbf r,\mathbf v,t)
 &=\mathbf F_-(\mathbf r,-m\gamma\mathbf v,t).
\end{aligned}
\label{eq:comparison_branch}
\end{equation}
Here $v=|\mathbf v|$, $\gamma$ is the corresponding Lorentz factor,
and $\nabla_v$ denotes the gradient with respect to velocity, with
$\partial_{v_i}=\partial/\partial v_i$.
The functions are expressed in velocity variables by substitution,
without a Jacobian factor. The corresponding relativistic transport
operator is
\begin{equation}
\begin{aligned}
 \mathcal D_e
 ={}&\partial_t+\mathbf v\cdot\nabla_r\\
 &-\frac{e}{m\gamma}
 \left(\delta_{ij}-\frac{v_i v_j}{c^2}\right)
 (\mathbf E+\mathbf v\times\mathbf B)_j
 \partial_{v_i},
\end{aligned}
\label{eq:comparison_transport}
\end{equation}
where, as before, repeated indices are summed over.

For the first comparison we keep only the leading non-relativistic spin
couplings, using $\gamma$=1. We also omit the spin-independent orbital gradient
corrections (of higher order in $\hbar$), to compare with the semi-classical orbital dynamics
of Ref.~\cite{Hurst2014}. Taking the non-relativistic limit alone
would not remove these corrections. Next, letting $A_{ij}$ reduce
to $\delta_{ij}$ , the single-species relations become
\begin{equation}
\begin{aligned}
 f_0&\simeq-f, & f_3&\simeq f,\\
 \mathbf g_0&\simeq\tfrac12\mathbf F,
 &\mathbf g_3&\simeq-\tfrac12\mathbf F.
\end{aligned}
\label{eq:comparison_nr_inverse}
\end{equation}
In this part of the calculation, the original DHW functions are
understood to be evaluated at $\mathbf p=-m\mathbf v$, so that
$\partial_{p_j}=-m^{-1}\partial_{v_j}$. The local spatial operator
therefore takes the form
\begin{equation}
 \mathbf D_e
 =\nabla_r-\frac{e}{m}\mathbf B\times\nabla_v.
\label{eq:comparison_spatial_operator}
\end{equation}

The leading non-relativistic  spin contribution to $\delta\mathbf g_1$ is
$\hbar\mathbf D_e\times\mathbf g_3/(2mc)$.
Using $\nabla_r\cdot\mathbf B=0$, we find the identities
\begin{equation}
\begin{aligned}
 \mathbf D_e\cdot(\mathbf D_e\times\mathbf g)
 &=-\frac{e}{m}
   (\partial_{r_j}B_i)\partial_{v_j}g_i,\\
 \mathbf P\cdot(\mathbf P\times\mathbf g)
 &=\frac{e\hbar^2}{4m}
   (\partial_{r_j}B_i)\partial_{v_j}g_i\\
 &\quad+\mathcal O(\hbar^4\mathbf g).
\end{aligned}
\label{eq:comparison_scalar_identities}
\end{equation}
In the second identity, $\mathbf P$ denotes the original momentum
operator expressed in these velocity variables. We keep its first
nonlocal correction, including momentum derivatives acting on the
explicit momentum factor. The spin-dependent contributions to
$Q_0$ and $Q_3$ are then
\begin{equation}
\begin{aligned}
 Q_0^{\mathrm{spin}}&\simeq
 \frac{e\hbar}{2m^2}
 (\partial_{r_j}B_i)\partial_{v_j}g_{3i},\\
 Q_3^{\mathrm{spin}}&\simeq
 \frac{e\hbar}{2m^2}
 (\partial_{r_j}B_i)\partial_{v_j}g_{0i}.
\end{aligned}
\label{eq:comparison_scalar_sources}
\end{equation}
Combining these terms in accordance with
Eq.~\eqref{eq:scalar_fpm_evolution} gives
\begin{equation}
 \mathcal D_e^{\mathrm{nr}}f
 =\frac{e\hbar}{4m^2}
   (\partial_{r_j}B_i)\partial_{v_j}F_i,
\label{eq:comparison_nr_scalar}
\end{equation}
where
\begin{equation}
\begin{aligned}
 \mathcal D_e^{\mathrm{nr}}
 ={}&\partial_t+\mathbf v\cdot\nabla_r\\
 &-\frac{e}{m}
   (\mathbf E+\mathbf v\times\mathbf B)\cdot\nabla_v.
\end{aligned}
\label{eq:comparison_nr_transport}
\end{equation}

For the vector equation we use
$\mathbf U\simeq c\mathbf D_e f_0$. The identity
\begin{equation}
 \mathbf D_e\times(\mathbf D_e f_0)
 =-\frac{e}{m}
   (\partial_{r_j}\mathbf B)\partial_{v_j}f_0
\label{eq:comparison_dcurl}
\end{equation}
shows that both vector sources at leading order contribute as:
\begin{equation}
\begin{aligned}
 \mathbf S_0&\simeq\frac{e\hbar}{2m^2}
   (\partial_{r_j}\mathbf B)\partial_{v_j}f_3,\\
 \mathbf S_3&\simeq\frac{e\hbar}{2m^2}
   (\partial_{r_j}\mathbf B)\partial_{v_j}f_0.
\end{aligned}
\label{eq:comparison_nr_vector_sources}
\end{equation}
The term proportional to $\mathbf p\times \mathbf U_t$  and the electric terms contained in
$\mathbf U$, in the expressions for $\mathbf{S}_0$ and $\mathbf{S}_3$, are relativistic corrections and have been omitted here.
Combining the two sources according to
Eq.~\eqref{eq:spin_Fpm_source} gives
\begin{equation}
\begin{aligned}
 \mathcal D_e^{\mathrm{nr}}\mathbf F
 -\frac{e}{m}\mathbf B\times\mathbf F
 =\frac{e\hbar}{m^2}
   (\partial_{r_j}\mathbf B)\partial_{v_j}f.
\end{aligned}
\label{eq:comparison_nr_vector}
\end{equation}

To compare the numerical coefficients, we denote the scalar and spin
functions of Ref.~\cite{Hurst2014} by $h_0$ and $\mathbf h$.
Their normalization relative to our functions is
\begin{equation}
 h_0=Cf,\qquad
 \mathbf h=C\frac{\hbar}{4}\mathbf F,
\label{eq:comparison_hurst_normalization}
\end{equation}
where $C$ accounts for the common phase-space normalization.
This relation follows since $\mathbf g_0=\mathbf F/2$ for a single
non-relativistic species, and the spin density includes a factor
$\hbar/2$.
The resulting equations are
\begin{align}
 \mathcal D_e^{\mathrm{nr}}h_0
 &-\frac{e}{m^2}
   (\partial_{r_j}B_i)\partial_{v_j}h_i=0,
 \label{eq:comparison_hurst_scalar}\\
 \mathcal D_e^{\mathrm{nr}}h_i
 &+\frac{e}{m}(\mathbf h\times\mathbf B)_i
 \nonumber\\
 &-\frac{\mu_B\hbar}{2m}
   (\partial_{r_j}B_i)\partial_{v_j}h_0=0,
 \label{eq:comparison_hurst_vector}
\end{align}
where $\mu_B=e\hbar/(2m)$ is the Bohr magneton. With the approximations stated above,
these magnetic-gradient terms and the precession term agree with
Eqs.~(12) and (13) of Ref.~\cite{Hurst2014}.

Next we keep the full relativistic velocity dependence and examine
the spin precession. Our vector equation becomes
\begin{equation}
\begin{aligned}
 \mathcal D_e\mathbf F
 &-\frac{e}{m\gamma}\mathbf B\times\mathbf F\\
 &+\frac{e}{mc^2\gamma}
   \mathbf v\times(\mathbf E\times\mathbf F)
 =\boldsymbol{\mathcal S}.
\end{aligned}
\label{eq:comparison_F_relativistic}
\end{equation}
Here $\boldsymbol{\mathcal S}$ denotes the electron-branch source
$\boldsymbol{\mathcal S}_-$, evaluated at $\mathbf p=-m\gamma\mathbf v$
and expressed in velocity variables by the same substitution as
$\mathbf F$.
The electric term is not resulting in a pure rotation of $\mathbf F$.
As we will see, this is due to the choice of spin variable, and nevertheless the result is consistent with Thomas precession.

To see this, we first rescale the spin function according to
\begin{equation}
 \mathbf a=\gamma\mathbf F,\qquad
 a^0=\frac{\mathbf v\cdot\mathbf a}{c}.
\label{eq:comparison_spin_fourvector}
\end{equation}
Along the classical electron trajectories we have
\begin{equation}
\begin{aligned}
 \mathcal D_e\mathbf v
 &=-\frac{e}{m\gamma}
 \left[
 \mathbf E+\mathbf v\times\mathbf B
 -\frac{\mathbf v(\mathbf E\cdot\mathbf v)}{c^2}
 \right],\\
 \mathcal D_e\gamma
 &=-\frac{e}{mc^2}\mathbf E\cdot\mathbf v.
\end{aligned}
\label{eq:comparison_characteristics}
\end{equation}
The derivative of $\gamma$ cancels the term proportional to
$-\mathbf F(\mathbf v\cdot\mathbf E)$ in the double cross product.
Thus the rescaled equation becomes
\begin{equation}
 \mathcal D_e\mathbf a
 =-\frac{e}{m\gamma}
   \left(\mathbf a\times\mathbf B
         +\frac{a^0}{c}\mathbf E\right)
   +\gamma\boldsymbol{\mathcal S}.
\label{eq:comparison_spatial_BMT}
\end{equation}
If the source is omitted, the temporal component satisfies
\begin{equation}
 \mathcal D_e a^0
 =-\frac{e}{mc\gamma}\mathbf E\cdot\mathbf a.
\label{eq:comparison_temporal_BMT}
\end{equation}
These are the spatial and temporal components of the BMT equation
for electrons with the spin $g$ factor set to $g=2$~\cite{BMT1959}, as expected.
With the metric signature $(+,-,-,-)$, the four-vector
$a^\mu=(a^0,\mathbf a)$ obeys $a^\mu U_\mu=0$, where
$U^\mu=\gamma(c,\mathbf v)$ is the particle four-velocity.
Greek indices run over the four space-time components, and repeated
Greek indices are contracted using this metric.
Consequently, apart from a constant normalization,
$\gamma\mathbf F$ evolves as the spatial part of the spin
four-vector when no source term is present.

To express the result in the particle rest frame, we introduce the
spin function $\boldsymbol\zeta$, with the same constant normalization.
The Lorentz transformation and its inverse give
\begin{equation}
\begin{aligned}
 a_i&=\zeta_i+
 \frac{\gamma^2 v_i v_j\zeta_j}{c^2(\gamma+1)},\\
 \zeta_i&=\gamma F_i-
 \frac{\gamma^2 v_i v_jF_j}{c^2(\gamma+1)}.
\end{aligned}
\label{eq:comparison_rest_frame}
\end{equation}
Differentiating these relations along the particle trajectories gives
\begin{equation}
\begin{aligned}
 \mathcal D_e\zeta_i
 &-\frac{e}{m}
 \left\{
 \left[
 \frac{\mathbf B}{\gamma}
 -\frac{\mathbf v\times\mathbf E}{c^2(\gamma+1)}
 \right]\times\boldsymbol\zeta
 \right\}_i\\
 &=\left(
 \gamma\delta_{ij}
 -\frac{\gamma^2 v_i v_j}{c^2(\gamma+1)}
 \right)\mathcal S_j.
\end{aligned}
\label{eq:comparison_rest_BMT}
\end{equation}
In particular, differentiating the coefficient in the inverse boost
gives
\begin{equation}
\begin{aligned}
 &\mathcal D_e\left(
 \gamma\delta_{ij}
 -\frac{\gamma^2 v_i v_j}{c^2(\gamma+1)}
 \right)\\
 &\quad=(\mathcal D_e\gamma)\delta_{ij}\\
 &\qquad-\frac{\gamma^2}{c^2(\gamma+1)}
 \left[
 (\mathcal D_e v_i)v_j+v_i(\mathcal D_e v_j)
 \right]\\
 &\qquad-\frac{\gamma(\gamma+2)v_i v_j}
 {c^2(\gamma+1)^2}\mathcal D_e\gamma.
\end{aligned}
\label{eq:comparison_K_derivative}
\end{equation}
Without including the coupling to the scalar source terms in the right-hand side of  Eq.~\eqref{eq:comparison_F_relativistic}, we thereby find agreement with the usual rest-frame BMT precession, which preserves $|\boldsymbol\zeta|$. Equivalently, the combination
\begin{equation}
 |\boldsymbol\zeta|^2
 =\gamma^2\left[
 |\mathbf F|^2-\frac{(\mathbf v\cdot\mathbf F)^2}{c^2}
 \right]
\label{eq:comparison_spin_invariant}
\end{equation}
is constant along a source-free trajectory.

In the weakly relativistic limit, the electric precession term
on the left-hand side becomes
\begin{equation}
 \frac{e}{2mc^2}
 (\mathbf v\times\mathbf E)\times\boldsymbol\zeta.
\label{eq:comparison_Thomas_limit}
\end{equation}
Thus we recover the usual Thomas factor $1/2$.
Specifically, the electric term agrees with Eq.~(4.5) of
Hurst \textit{et al.}~\cite{Hurst2017}, where the effective magnetic
field is $\mathbf B-\mathbf v\times\mathbf E/(2c^2)$.
The comparison here establishes agreement of the Thomas
term. The rescaling and Lorentz transformation above explain why our original
precession equation for $\mathbf F$ has a different appearance.

\section{Expansion to arbitrary order and inclusion of vacuum polarization}
\label{sec:all_orders_vacuum}

In the preceding section, the scalar equations were derived up to
second order in the weak-quantum expansion, whereas the vector equations
kept the leading spin and magnetization contributions. The same
reduction can be extended systematically by improving the solutions
for $f_1$, $f_2$, $\mathbf g_1$, and $\mathbf g_2$ through successive
substitutions. Similarly, the nonlocal operators can be expanded to any
order. Throughout, weakly relativistic approximations are avoided, and the full dependence on $p/(mc)$ is kept.
The expansion assumes electromagnetic fields that are weak compared to the
critical fields, as well as temporal and spatial variations that are slow compared to
the Compton scales. It also selects the slowly varying branch of the
DHW dynamics, excluding rapid oscillations associated with
the eliminated variables. Under these assumptions, one obtains a
formal reduction to eight independent phase-space functions that can be performed to all orders. Note, however, that exponentially small Schwinger pair production, will not be recovered
by a finite truncation. Perturbative vacuum polarization, however, will be covered by the reduced description.

\subsection{The expansion to all orders}
\label{subsec:all_order_elimination}

We introduce the characteristic time constant $\tau=\hbar/(2mc^2)$, which is equal to one-half of the inverse Compton frequency. In this section, we use the full operators $D_t$, $\mathbf D$, and $\mathbf P$.
Their order is kept as written; in particular, we do not commute $D_t$
with $\mathbf D$ or $\mathbf P$. In order to eliminate $f_1$ and $f_2$, it is useful to define the auxiliary scalar
\begin{equation}
 L=D_t\left(\frac{\mathbf P\cdot\mathbf g_3}{mc}\right)
                      +c\mathbf D\cdot\mathbf g_0.
\label{eq:ao_scalar_seeds}
\end{equation}
Eliminating $f_1$ and $f_2$ from
Eqs.~\eqref{eq:spin_f1_exact} and \eqref{eq:spin_f2_exact} gives
\begin{equation}
\begin{aligned}
 (1+\tau^2D_t^2)f_1
 &=\frac{\mathbf P\cdot\mathbf g_3}{mc}
          -\tau^2cD_t(\mathbf D\cdot\mathbf g_0),\\
 (1+\tau^2D_t^2)f_2&=-\tau L.
\end{aligned}
\label{eq:ao_scalar_second_order}
\end{equation}
For slow time dependence we expand the inverse operator as
\begin{equation}
 (1+\tau^2D_t^2)^{-1}
   =\sum_{n=0}^{\infty}(-\tau^2D_t^2)^n.
\label{eq:ao_inverse}
\end{equation}
The successive approximations are then
\begin{align}
 f_1={}&\frac{\mathbf P\cdot\mathbf g_3}{mc}
       -\tau^2D_tL+\tau^4D_t^3L\nonumber\\
       &-\tau^6D_t^5L+\cdots,\label{eq:ao_f1}\\
 f_2={}&-\tau L+\tau^3D_t^2L-\tau^5D_t^4L+\cdots.
 \label{eq:ao_f2}
\end{align}

The terms displayed for $f_1$ start at orders $\hbar$, $\hbar^3$,
$\hbar^5$, and $\hbar^7$, while those for $f_2$ start at
$\hbar^2$, $\hbar^4$, and $\hbar^6$. Since the full nonlocal
operators are still present, each term also contains higher-order
spatial-gradient corrections. These could be expanded and collected
in case a particular order in $\hbar$ is what we aim for.

Similarly, in order to calculate expressions for $\mathbf g_1$ and 
$\mathbf g_2$, we define the auxiliary vectors $\mathbf V$ and $\mathbf R$ by
\begin{equation}
\begin{aligned}
 \mathbf V&=\frac{\mathbf P\times\mathbf g_0}{mc}
 -\tau\left[D_t\left(\frac{\mathbf P f_3}{mc}\right)
                                +c\mathbf D f_0\right],\\
 \mathbf R&=\tau(D_t\mathbf V+c\mathbf D\times\mathbf g_3).
\end{aligned}
\label{eq:ao_vector_seeds}
\end{equation}
The exact vector relations can now be written as
\begin{equation}
\begin{aligned}
 \mathbf g_2&=\mathbf V-\tau D_t\Big(\mathbf g_1-\frac{\mathbf P f_3}{mc}\Big),\\
 \mathbf g_1-\frac{\mathbf P f_3}{mc}&=\mathbf R
               -\tau^2D_t^2\Big(\mathbf g_1-\frac{\mathbf P f_3}{mc}\Big).
\end{aligned}
\label{eq:ao_vector_relations}
\end{equation}
Expanding the same inverse operator gives
\begin{align}
 \mathbf g_1={}&\frac{\mathbf P f_3}{mc}+\mathbf R
     -\tau^2D_t^2\mathbf R
     +\tau^4D_t^4\mathbf R+\cdots,
 \label{eq:ao_g1}\\
 \mathbf g_2={}&\mathbf V-\tau D_t\mathbf R
     +\tau^3D_t^3\mathbf R
     -\tau^5D_t^5\mathbf R+\cdots.
 \label{eq:ao_g2}
\end{align}
Here $\mathbf V$ is of order $\hbar$, and $\mathbf R$ is of order
$\hbar^2$. Thus the terms in $\mathbf g_1$ start at orders $1$,
$\hbar^2$, $\hbar^4$, and $\hbar^6$, and those in $\mathbf g_2$
at $\hbar$, $\hbar^3$, $\hbar^5$, and $\hbar^7$.
Note that the first nonlocal correction in $\mathbf P f_3/(mc)$ is
already of order $\hbar^2$, and would be combined with the leading
contribution from $\mathbf R$, if we treated all expansion parameters in $\hbar$ on the same footing.

For a given truncation, we expand the operators and collect the terms
according to the ordering in Sec.~\ref{sec:scalar_evolution}.
Inserting the resulting expressions into the remaining DHW equations
gives higher-order versions of the scalar and vector equations derived
above. As before, factors $1/\hbar$ in the original equations must
be taken into account when deciding how far to carry the substitution.

Time derivatives of functions inside correction terms are replaced successively using the lower-order evolution equations.
For example, in the second-order scalar corrections we use the
classical equations for $f_0$ and $f_3$. Repeating this procedure at
higher orders leaves first-order evolution equations for eight
independent functions. Time derivatives
of prescribed fields may remain; for self-consistent fields these can
be related to Maxwell's equations. The variables $f_\pm$ and
$\mathbf F_\pm$ are still used to approximately separate electron and positron degrees of freedom.

\subsection{Vacuum contributions and their electromagnetic response}
\label{subsec:ao_vacuum_response}

Nothing in the elimination scheme described in the previous subsection requires us to subtract the vacuum contribution to the phase-space variables.
Using the free-vacuum values and the definition of $f_\pm$, we have
\begin{equation}
\begin{aligned}
 f_{3,\mathrm{vac}}&=-\frac{2}{\gamma_p},&
 \mathbf g_{1,\mathrm{vac}}&=-\frac{2\mathbf p}{mc\gamma_p},\\
 f_{0,\mathrm{vac}}&=0,& f_{\pm,\mathrm{vac}}&=-1.
\end{aligned}
\label{eq:ao_vacuum_branches}
\end{equation}
Thus $f_\pm$ includes a vacuum contribution, and should not in general be interpreted
as a classical distribution function. Specifically, a change in these functions due to an applied
field does not need to imply that real particles have been created.

Although the free vacuum is homogeneous and stationary, its original
DHW functions depend on momentum. For example, the local electric-field term gives
\begin{equation}
 D_{t,\mathrm{loc}} f_{3,\mathrm{vac}}
    =e\mathbf E\cdot\nabla_p f_{3,\mathrm{vac}}.
\label{eq:ao_vacuum_forcing}
\end{equation}

The nonlocal corrections can also act on the unperturbed vacuum:
their spatial derivatives act on the fields, whereas their momentum
derivatives act on the vacuum functions. Similarly, derivatives in $\mathbf D$ can act on the
vacuum functions, including nonlocal terms when field gradients are
present, although some contractions vanish by symmetry. By contrast,
all derivatives acting directly on the constants $f_{\pm,\mathrm{vac}}=-1$
are zero. In the $f_\pm$ representation, the response instead arises
through momentum-dependent coefficients and the reconstructed DHW
functions. Once fields are applied, the vacuum corrections generally
vary in both space and time.

To identify the expected integrated response, we first consider the
weak-field Euler--Heisenberg (EH) effective Lagrangian density (see, e.g., \cite{MarklundShukla2006,eriksson2004possibility})
$\mathcal L_{\mathrm{EH}}$. Omitting field derivatives, it is
\begin{equation}
\begin{aligned}
 \mathcal L_{\mathrm{EH}}
 ={}&\frac{\epsilon_0}{2}(E^2-c^2B^2)\\
 &+\kappa\left[(E^2-c^2B^2)^2
              +7c^2(\mathbf E\cdot\mathbf B)^2\right],\\
 \kappa={}&\frac{\alpha\epsilon_0}{90\pi E_{\mathrm{c}}^2},\qquad
 E_{\mathrm{c}}=\frac{m^2c^3}{e\hbar},
\end{aligned}
\label{eq:ao_EH}
\end{equation}
where, in this subsection, we let $e>0$ denote the elementary charge (rather than the electron charge),  
$\epsilon_0$ is the vacuum permittivity, $\alpha$ is the fine-structure
constant, and $E_{\mathrm{c}}$ is the Schwinger critical electric field.
We use $E=|\mathbf E|$ and $B=|\mathbf B|$. The first term is the Maxwell Lagrangian,
and the remaining terms are the leading one-loop Euler--Heisenberg
correction. The corresponding polarization and magnetization expressions are
\begin{align}
 \mathbf P_{\mathrm{EH}}
 ={}&4\kappa(E^2-c^2B^2)\mathbf E\nonumber\\
 &+14\kappa c^2(\mathbf E\cdot\mathbf B)\mathbf B,
 \label{eq:ao_EH_P}\\
 \mathbf M_{\mathrm{EH}}
 ={}&-4\kappa c^2(E^2-c^2B^2)\mathbf B\nonumber\\
 &+14\kappa c^2(\mathbf E\cdot\mathbf B)\mathbf E,
 \label{eq:ao_EH_M}
\end{align}
giving the charge and current densities
\begin{equation}
\begin{aligned}
 \rho_{\mathrm{EH}}&=-\nabla_r\cdot\mathbf P_{\mathrm{EH}},\\
 \mathbf j_{\mathrm{EH}}&=\partial_t\mathbf P_{\mathrm{EH}}
               +\nabla_r\times\mathbf M_{\mathrm{EH}}.
\end{aligned}
\label{eq:ao_EH_sources}
\end{equation}

The correction to the Lagrangian is fourth order in field amplitude.
Consequently, the polarization and magnetization, and the associated
sources, are third order in the fields.

There is also a finite contribution that is linear in field amplitude,
but contains additional space-time derivatives. For general electric
and magnetic fields, the leading derivative contribution to the
renormalized vacuum response can be written as
\begin{equation}
\begin{aligned}
 \rho_D&=\frac{\alpha\epsilon_0c^2}{15\pi\omega_c^2}
       \Box(\nabla_r\cdot\mathbf E),\\
 \mathbf j_D&=-\frac{\alpha\epsilon_0c^2}{15\pi\omega_c^2}
       \Box\bigl(\partial_t\mathbf E
                    -c^2\nabla_r\times\mathbf B\bigr),\\
 \Box&=\frac{1}{c^2}\partial_t^2-\nabla_r^2,
 \qquad \omega_c=\frac{mc^2}{\hbar}.
\end{aligned}
\label{eq:ao_derivative_current}
\end{equation}
This is the leading weak-field, low-frequency and long-wavelength
term; higher derivative and nonlinear derivative corrections are
omitted. Here $\Box$ is the d'Alembert operator, $\omega_c$ is the
Compton angular frequency, and the subscript $D$ labels the derivative
contribution to the charge and current densities.
The electromagnetic derivative correction is discussed in
Sec.~II D of Ref.~\cite{MarklundShukla2006}. The coefficient used here
is the standard renormalized one-loop coefficient, consistent with
the quadratic weak-field term in Eq.~(20) of
Ref.~\cite{Karbstein2021Derivative}.

In our case, to obtain the vacuum response from the reduced equations, we must keep
the induced corrections to the eight independent functions as well
as the reconstructed auxiliary functions. The physical charge and
current densities are given by
Eqs.~\eqref{eq:DHW_charge_density} and \eqref{eq:DHW_current_density}.
Although the free-vacuum current integrates to zero, the induced
current density does not. 

We will not carry out the above procedure in full detail. However, this was done for the special case of longitudinal electrostatic fields
by Lundstr\"om \textit{et al.}\ \cite{lundstrom2026modified}, using the same approach as here (i.e. expanding the DHW-equations) exactly recovering the cubic vacuum polarization and the derivative corrections. The key features of that calculation also apply here. First, the
integrated linear response contains a logarithmic ultraviolet
divergence, which is absorbed into the renormalized charge; see
Sec.~III C of that work. For general electromagnetic fields the
component equations are more involved, but the divergent term still
renormalizes the Maxwell action. The regularization and subtraction
must preserve gauge invariance and current conservation, with the
physical charge fixed by the same low-energy condition.

Secondly, some apparently low-order vacuum contributions vanish after
momentum integration. This will follow from momentum parity, total
momentum derivatives, or cancellations between the induced distributions
and the reconstructed current. The same regulator must be used
throughout these cancellations. In particular, surface terms from
divergent integrals cannot be dropped before regularization.

Thirdly, real particles modify the quantum response, in addition to
the usual vacuum contribution. Equation (68) of
Ref.~\cite{lundstrom2026modified} separates these contributions, and
Eq.~(69) in that work gives the result after taking a further ultrarelativistic
limit. The modification is consistent with Pauli blocking in occupied
momentum regions. However, it also depends on the particle distribution
and its evolution, and cannot in general be represented by a single
multiplicative blocking factor.

The electrostatic calculation therefore shows explicitly that
perturbative elimination can cover both the effect of vacuum polarization and
quantum corrections involving real particles. The substitutions in
Eqs.~\eqref{eq:ao_f1}, \eqref{eq:ao_f2}, \eqref{eq:ao_g1}, and
\eqref{eq:ao_g2} extend the construction to general fields.
Naturally, evaluating the full electromagnetic vacuum response also requires
momentum integration and renormalization. In this approach, vacuum polarization follows from the coupling between the electron and positron degrees of freedom, when keeping coupling terms to the required order.
\section{Summary and Conclusion}
\label{Summary}
In this paper, we have studied expansions of the DHW equations in
weak fields (compared to the critical field) and on spatial and temporal scales long compared with the corresponding
Compton scales. This generalizes a previous work, Ref.~\cite{lundstrom2026modified}, to account for general electromagnetic fields and a full 3D geometry. Similar expansion schemes may allow
for a FW transformation (see, e.g.,
Refs.~\cite{silenko2015general,hinschberger2016coherent,ekman2017relativistic}),
separating the electron and positron degrees of freedom. Here,
however, we have deliberately avoided a full separation of the Dirac
field into electrons and positrons. From a philosophical point of
view, the separation of electrons and positrons is only approximate
in the Dirac theory, and residual coupling need not necessarily be
avoided. We refer to the approximate equations keeping the electron-positron coupling as the WCEP equations. A clear demonstration of the usefulness of including this
coupling is that it allows vacuum polarization to be covered by
the reduced model. Often, when plasmas in strong astrophysical fields
are studied, currents due to vacuum polarization are added to the
free plasma currents phenomenologically (see, e.g., Refs.~\cite{huang2015quantum,medvedev2023plasma,chen2013nonlinear}). This may be a useful
approximation when the plasma density is comparatively low. 
The relative importance of matter-dependent electron--positron coupling and vacuum polarization depends on the plasma parameters and field configuration. This question has been discussed in some detail for the electrostatic case in Ref.~\cite{lundstrom2026modified}.
From the approach presented here, it is apparent that vacuum
polarization is due to the mixing of electron and positron states.
Since the coupling of virtual states and real-particle states is
described by the same equations, a more fundamental and consistent
approach should include vacuum polarization and real-particle
electron--positron coupling on an equal footing.

Even for plasmas where vacuum polarization can be neglected,
describing the system in terms of the variables $f_\pm$ and
$\mathbf F_\pm$ potentially has advantages, but also some drawbacks.
The most obvious drawback is that, when high accuracy is required,
we need to keep all eight components (one scalar equation and one vector equation) for both electrons and positrons, even if only one species is present in the plasma of
interest. In such a case, a related approach based on the FW
transformation would require only four variables. To some extent,
this difference is due to the present approach capturing
additional physics, in particular the vacuum polarization discussed above.
However, much of the physics captured using the FW transformation
would, in our case, be encoded in the eight equations for coupled
electrons and positrons, rather than just four equations for a single
species. Keeping eight rather than four independent functions increases the size
of the system, but simplifies the relation between momentum and velocity. Specifically, in a fully relativistic
theory, complete separation of electrons and positrons comes at a
certain cost: a very complicated relation between momentum and
kinetic velocity. This is illustrated by Eq.~(26) of
Ref.~\cite{ekman2017relativistic}, and also by the current source
in Eq.~(20) of that work, which requires the kinetic velocity to
calculate the free current density. Moreover, if higher accuracy
had been sought in Ref.~\cite{ekman2017relativistic}, the complicated
expression for the velocity would also have entered into the expression
for the Lorentz force.

In our work, we have given fairly limited attention to the source
terms in Eqs.~\eqref{eq:scalar_fpm_evolution} and
\eqref{eq:spin_Fpm_evolution} that couple electrons and positrons.
The reason is that there are several possible ways to write these
terms for various types of plasmas. Except for certain specific comparisons,
we have treated all expansion parameters involving $\hbar$ on the
same footing. However, for particular plasma parameters and field
configurations, the ordering of the possible expansion parameters
could differ considerably. For simplicity, let $\omega$ denote a
characteristic frequency of the field variations and $k$ a
characteristic wave number. 
For field variations satisfying the Faraday-law estimate $\omega B\sim kE$, the parameter $\hbar k^2/m\omega$ estimates the strength of the magnetic dipole force relative to the electric part of the Lorentz force for a particle in a given spin state.
If this parameter is much larger than
$\hbar k/p_{\mathrm{ch}}$, where $p_{\mathrm{ch}}$ is a characteristic
particle momentum, it would make sense to use a scheme that keeps the spin-dependent
contributions at the order considered, including the magnetic-gradient
coupling, while omitting the orbital dispersive corrections. This is
the approximation used to obtain the coupling term in
Eq.~\eqref{eq:scalar_fpm_local_source}. However, if the dimensionless
parameters were of the same order, an expansion to a given order in
$\hbar$ would make sense. Other dimensionless parameters, such as
$\mu_B B/(k_B T)$ and $\hbar\omega_p/(k_B T)$, could also enter in a
more complete picture. Here $k_B$ is Boltzmann's constant, $T$ is
the temperature, and $\omega_p$ is the plasma frequency.
See Refs.~\cite{brodin2022quantum,lundin2010linearized} for more
complete discussions of the role of different dimensionless quantum
parameters.

The main conclusion of the present work is that the expansion
achieves the goal of generating weakly coupled electron and positron
distributions, provided the basic assumptions of weak fields
(compared to the critical field) and slow scales (compared to
Compton scales) are fulfilled. Besides the expansion procedure
itself, the key ingredient is the change to the new dependent
variables $\mathbf F_\pm$ and $f_\pm$. Specifically, using a spin-like function that is a momentum-dependent linear combination of the original spin and magnetization is necessary to achieve the
approximate separation. If we additionally set the linear combination to have unity in front of $\mathbf g_0$, the separation condition fixes the momentum dependence in front of $\mathbf g_3$. Thus, in a regime where the expansion applies, the choice of $\mathbf F_\pm$ as a suitable variable is naturally imposed by the DHW equations themselves.

Applications of the approach developed in this work include at least two different aspects. Firstly, the present work can serve as a starting point for development of simplified models. The fact that all quantum relativistic effects that can be captured in a perturbative framework are covered is the main advantage. Secondly, the model developed here offers a unified treatment in strong-field environments \cite{brodin2022quantum,lundstrom2026modified,zhang2020relativistic,medvedev2023plasma}, when both vacuum polarization and real particle quantum effects are of importance.   

\bibliographystyle{apsrev4-2}
\bibliography{References}

@article{melrose2020quantum,
  author = {Melrose, D. B.},
  title = {Quantum kinetic theory for unmagnetized and magnetized plasmas: A tutorial review of quantum plasma theory},
  journal = {Rev. Mod. Plasma Phys.},
  volume = {4},
  number = {1},
  pages = {8},
  year = {2020},
}

@article{brodin2022quantum,
  author = {Brodin, Gert and Zamanian, Jens},
  title = {Quantum kinetic theory of plasmas},
  journal = {Rev. Mod. Plasma Phys.},
  volume = {6},
  pages = {4},
  year = {2022},
  doi = {10.1007/s41614-022-00065-5},
}

@article{manfredi2019phase,
  author = {Manfredi, Giovanni and Hervieux, Paul-Antoine and Hurst, J{\'e}r{\^o}me},
  title = {Phase-space modeling of solid-state plasmas: A journey from classical to quantum},
  journal = {Rev. Mod. Plasma Phys.},
  volume = {3},
  number = {1},
  pages = {13},
  year = {2019},
}

@article{manfredi2021fluid,
  author = {Manfredi, Giovanni and Hervieux, Paul-Antoine and Hurst, J{\'e}r{\^o}me},
  title = {Fluid descriptions of quantum plasmas},
  journal = {Rev. Mod. Plasma Phys.},
  volume = {5},
  number = {1},
  pages = {7},
  year = {2021},
}

@book{bonitz2016quantum,
  author = {Bonitz, Michael},
  title = {Quantum Kinetic Theory},
  edition = {2},
  publisher = {Springer},
  address = {Cham},
  year = {2016},
  doi = {10.1007/978-3-319-24121-0},
}

@article{Hurst2014,
  author = {Hurst, J{\'e}r{\^o}me and Morandi, Omar and Manfredi, Giovanni and Hervieux, Paul-Antoine},
  title = {Semiclassical {Vlasov} and fluid models for an electron gas with spin effects},
  journal = {Eur. Phys. J. D},
  volume = {68},
  pages = {176},
  year = {2014},
  doi = {10.1140/epjd/e2014-50205-5},
}

@article{asenjo2011hydrodynamical,
  author = {Asenjo, Felipe A and Mu{\~n}oz, V{\'\i}ctor and Valdivia, J Alejandro and Mahajan, Swadesh M},
  title = {A hydrodynamical model for relativistic spin quantum plasmas},
  journal = {Phys. Plasmas},
  volume = {18},
  number = {1},
  pages = {012107},
  year = {2011},
  doi = {10.1063/1.3533448},
}

@article{andreev2015separated,
  author = {Andreev, Pavel A},
  title = {Separated spin-up and spin-down quantum hydrodynamics of degenerated electrons: Spin-electron acoustic wave appearance},
  journal = {Phys. Rev. E},
  volume = {91},
  number = {3},
  pages = {033111},
  year = {2015},
  doi = {10.1103/PhysRevE.91.033111},
}

@article{haas2010fluid,
  author = {Haas, F. and Marklund, M. and Brodin, G. and Zamanian, J.},
  title = {Fluid moment hierarchy equations derived from quantum kinetic theory},
  journal = {Phys. Lett. A},
  volume = {374},
  number = {3},
  pages = {481--484},
  year = {2010},
}

@article{misra2010generation,
  author = {Misra, A. P. and Brodin, G. and Marklund, M. and Shukla, P. K.},
  title = {Generation of wakefields by whistlers in spin quantum magnetoplasmas},
  journal = {Phys. Plasmas},
  volume = {17},
  number = {12},
  pages = {122306},
  year = {2010},
  doi = {10.1063/1.3527995},
}

@article{zamanian2010scalar,
  author = {Zamanian, Jens and Marklund, Mattias and Brodin, Gert},
  title = {Scalar quantum kinetic theory for spin-1/2 particles: mean field theory},
  journal = {New J. Phys.},
  volume = {12},
  number = {4},
  pages = {043019},
  year = {2010},
}

@article{brodin2008effects,
  author = {Brodin, Gert and Marklund, Mattias and Zamanian, Jens and Ericsson, {\AA}sa and Mana, Piero L},
  title = {Effects of the g factor in semiclassical kinetic plasma theory},
  journal = {Phys. Rev. Lett.},
  volume = {101},
  number = {24},
  pages = {245002},
  year = {2008},
  doi = {10.1103/PhysRevLett.101.245002},
}

@article{bakri2026wave,
  author = {Bakri, Benjamin and Crouseilles, Nicolas and Hervieux, Paul-Antoine and Manfredi, Giovanni},
  title = {Wave--particle interactions in a spin polarized plasma},
  journal = {Phys. Plasmas},
  volume = {33},
  number = {2},
  pages = {022111},
  year = {2026},
  doi = {10.1063/5.0312295},
}

@article{del2017spin,
  author = {Del Sorbo, Dario and Seipt, Daniel and Blackburn, Tom G and Thomas, Alexander G. R. and Murphy, Christopher D and Kirk, John G and Ridgers, C. P.},
  title = {Spin polarization of electrons by ultraintense lasers},
  journal = {Phys. Rev. A},
  volume = {96},
  number = {4},
  pages = {043407},
  year = {2017},
  doi = {10.1103/PhysRevA.96.043407},
}

@article{asenjo2012semi,
  author = {Asenjo, Felipe A and Zamanian, Jens and Marklund, Mattias and Brodin, Gert and Johansson, Petter},
  title = {Semi-relativistic effects in spin-1/2 quantum plasmas},
  journal = {New J. Phys.},
  volume = {14},
  number = {7},
  pages = {073042},
  year = {2012},
}

@article{Hurst2017,
  author = {Hurst, J{\'e}r{\^o}me and Hervieux, Paul-Antoine and Manfredi, Giovanni},
  title = {Phase-space methods for the spin dynamics in condensed matter systems},
  journal = {Philos. Trans. R. Soc. A},
  volume = {375},
  number = {2092},
  pages = {20160199},
  year = {2017},
  doi = {10.1098/rsta.2016.0199},
}

@article{ekman2017relativistic,
  author = {Ekman, Robin and Asenjo, Felipe A and Zamanian, Jens},
  title = {Relativistic kinetic equation for spin-1/2 particles in the long-scale-length approximation},
  journal = {Phys. Rev. E},
  volume = {96},
  number = {2},
  pages = {023207},
  year = {2017},
  doi = {10.1103/PhysRevE.96.023207},
}

@article{BMT1959,
  author = {Bargmann, V. and Michel, L. and Telegdi, V. L.},
  title = {Precession of the polarization of particles moving in a homogeneous electromagnetic field},
  journal = {Phys. Rev. Lett.},
  volume = {2},
  number = {10},
  pages = {435--436},
  year = {1959},
  doi = {10.1103/PhysRevLett.2.435},
}

@article{vasak1987quantum,
  author = {Vasak, David and Gyulassy, Miklos and Elze, Hans-Thomas},
  title = {Quantum transport theory for {Abelian} plasmas},
  journal = {Ann. Phys.},
  volume = {173},
  number = {2},
  pages = {462--492},
  year = {1987},
}

@article{BialynickiBirula1991,
  author = {Bia{\l}ynicki-Birula, Iwo and G{\'o}rnicki, Pawe{\l} and Rafelski, Johann},
  title = {Phase-space structure of the Dirac vacuum},
  journal = {Phys. Rev. D},
  volume = {44},
  number = {6},
  pages = {1825},
  year = {1991},
  doi = {10.1103/PhysRevD.44.1825},
}

@article{li2021enhanced,
  author = {Li, Lie-Juan and Mohamedsedik, Melike and Xie, Bai-Song},
  title = {Enhanced dynamically assisted pair production in spatial inhomogeneous electric fields with the frequency chirping},
  journal = {Phys. Rev. D},
  volume = {104},
  number = {3},
  pages = {036015},
  year = {2021},
  doi = {10.1103/PhysRevD.104.036015},
}

@article{hebenstreit2011pair,
  author = {Hebenstreit, Florian and Ilderton, Anton and Marklund, Mattias},
  title = {Pair production: The view from the lightfront},
  journal = {Phys. Rev. D},
  volume = {84},
  number = {12},
  pages = {125022},
  year = {2011},
  doi = {10.1103/PhysRevD.84.125022},
}

@article{aleksandrov2020pair,
  author = {Aleksandrov, Ivan A and Kohlf{\"u}rst, Christian},
  title = {Pair production in temporally and spatially oscillating fields},
  journal = {Phys. Rev. D},
  volume = {101},
  number = {9},
  pages = {096009},
  year = {2020},
  doi = {10.1103/PhysRevD.101.096009},
}

@article{edwards2025relativistic,
  author = {Edwards, James P and Ahmadiniaz, Naser and Schmidt, Sebastian M and Kohlf{\"u}rst, Christian},
  title = {Relativistic quantum kinetic theory: Higher order contributions in assisted {Schwinger} pair production},
  journal = {Phys. Rev. D},
  volume = {112},
  number = {3},
  pages = {L031901},
  year = {2025},
  doi = {10.1103/PhysRevD.112.L031901},
}

@article{al2021plasma,
  author = {Al-Naseri, Haidar and Zamanian, Jens and Brodin, Gert},
  title = {Plasma dynamics and vacuum pair creation using the Dirac-Heisenberg-{Wigner} formalism},
  journal = {Phys. Rev. E},
  volume = {104},
  number = {1},
  pages = {015207},
  year = {2021},
  doi = {10.1103/PhysRevE.104.015207},
}

@article{brodin2023plasma,
  author = {Brodin, Gert and Al-Naseri, Haidar and Zamanian, Jens and Torgrimsson, Greger and Eliasson, Bengt},
  title = {Plasma dynamics at the {Schwinger} limit and beyond},
  journal = {Phys. Rev. E},
  volume = {107},
  number = {3},
  pages = {035204},
  year = {2023},
  doi = {10.1103/PhysRevE.107.035204},
}

@article{nematizadeh2021study,
  author = {Nematizadeh Juneghani, Safura and Bagheri, Mehran and Shokri, Babak},
  title = {Study Zitterbewegung effect in a quasi one-dimensional relativistic quantum plasma by Dirac-Heisenberg-{Wigner} formalization},
  journal = {J. Cosmol. Astropart. Phys.},
  volume = {2021},
  number = {09},
  pages = {002},
  year = {2021},
}

@article{morandi2011wigner,
  author = {Morandi, Omar and Sch{\"u}rrer, Ferdinand},
  title = {{Wigner} model for quantum transport in graphene},
  journal = {J. Phys. A: Math. Theor.},
  volume = {44},
  number = {26},
  pages = {265301},
  year = {2011},
  doi = {10.1088/1751-8113/44/26/265301},
}

@article{akal2016low,
  author = {Akal, Ibrahim and Egger, Reinhold and M{\"u}ller, Carsten and Villalba-Ch{\'a}vez, Selym},
  title = {Low-dimensional approach to pair production in an oscillating electric field: Application to bandgap graphene layers},
  journal = {Phys. Rev. D},
  volume = {93},
  pages = {116006},
  year = {2016},
  doi = {10.1103/PhysRevD.93.116006},
}

@article{silenko2015general,
  author = {Silenko, Alexander J},
  title = {General method of the relativistic {Foldy-Wouthuysen} transformation and proof of validity of the {Foldy-Wouthuysen} {Hamiltonian}},
  journal = {Phys. Rev. A},
  volume = {91},
  number = {2},
  pages = {022103},
  year = {2015},
  doi = {10.1103/PhysRevA.91.022103},
}

@article{hinschberger2016coherent,
  author = {Hinschberger, Yannick and Manfredi, Giovanni and Hervieux, Paul-Antoine},
  title = {Coherent spin-light-induced mechanisms in the semirelativistic limit of the self-consistent {Dirac-Maxwell} equations},
  journal = {Phys. Rev. A},
  volume = {93},
  number = {4},
  pages = {042117},
  year = {2016},
  doi = {10.1103/PhysRevA.93.042117},
}

@article{lundstrom2026modified,
  author = {Lundstr{\"o}m, Sebastian and Semr{\'e}n, Philip and Al-Naseri, Haidar and Brodin, Gert},
  title = {Modified vacuum polarization in the presence of a plasma},
  journal = {Phys. Plasmas},
  volume = {33},
  number = {3},
  pages = {032104},
  year = {2026},
  doi = {10.1063/5.0317285},
}

@book{jackson1999classical,
  author = {Jackson, John David},
  title = {Classical Electrodynamics},
  edition = {3},
  publisher = {Wiley},
  address = {New York},
  year = {1999},
}

@article{hebenstreit2010schwinger,
  author = {Hebenstreit, Florian and Alkofer, Reinhard and Gies, Holger},
  title = {{Schwinger} pair production in space- and time-dependent electric fields: Relating the {Wigner} formalism to quantum kinetic theory},
  journal = {Phys. Rev. D},
  volume = {82},
  number = {10},
  pages = {105026},
  year = {2010},
  doi = {10.1103/PhysRevD.82.105026},
}

@article{gonoskov2022charged,
  author = {Gonoskov, A and Blackburn, T. G. and Marklund, M and Bulanov, S. S.},
  title = {Charged particle motion and radiation in strong electromagnetic fields},
  journal = {Rev. Mod. Phys.},
  volume = {94},
  number = {4},
  pages = {045001},
  year = {2022},
}

@article{fedotov2023advances,
  author = {Fedotov, A. and Ilderton, A. and Karbstein, F. and King, B. and Seipt, D. and Taya, H. and Torgrimsson, G.},
  title = {Advances in {QED} with intense background fields},
  journal = {Phys. Rep.},
  volume = {1010},
  pages = {1--138},
  year = {2023},
  doi = {10.1016/j.physrep.2023.01.003},
}

@article{MarklundShukla2006,
  author = {Marklund, Mattias and Shukla, Padma K.},
  title = {Nonlinear collective effects in photon-photon and photon-plasma interactions},
  journal = {Rev. Mod. Phys.},
  volume = {78},
  pages = {591--640},
  year = {2006},
  doi = {10.1103/RevModPhys.78.591},
}

@article{eriksson2004possibility,
  author = {Eriksson, Daniel and Brodin, Gert and Marklund, Mattias and Stenflo, Lennart},
  title = {Possibility to measure elastic photon-photon scattering in vacuum},
  journal = {Phys. Rev. A},
  volume = {70},
  number = {1},
  pages = {013808},
  year = {2004},
  doi = {10.1103/PhysRevA.70.013808},
}

@article{Karbstein2021Derivative,
  author = {Karbstein, Felix},
  title = {Derivative corrections to the {Heisenberg-Euler} effective action},
  journal = {J. High Energy Phys.},
  volume = {2021},
  number = {9},
  pages = {070},
  year = {2021},
  doi = {10.1007/JHEP09(2021)070},
}

@article{huang2015quantum,
  author = {Huang, Y. S.},
  title = {Quantum-electrodynamical birefringence vanishing in a thermal relativistic pair plasma},
  journal = {Sci. Rep.},
  volume = {5},
  number = {1},
  pages = {15866},
  year = {2015},
}

@article{medvedev2023plasma,
  author = {Medvedev, Mikhail V},
  title = {Plasma modes in {QED} super-strong magnetic fields of magnetars and laser plasmas},
  journal = {Phys. Plasmas},
  volume = {30},
  number = {9},
  pages = {092112},
  year = {2023},
  doi = {10.1063/5.0160628},
}

@article{chen2013nonlinear,
  author = {Chen, Wenbo and Bu, Zhigang and Li, Hehe and Luo, Yuee and Ji, Peiyong},
  title = {Nonlinear theory of intense laser-plasma interactions modified by vacuum polarization effects},
  journal = {Phys. Plasmas},
  volume = {20},
  number = {7},
  pages = {072113},
  year = {2013},
  doi = {10.1063/1.4816704},
}

@article{lundin2010linearized,
  author = {Lundin, J. and Brodin, G.},
  title = {Linearized kinetic theory of spin-1/2 particles in magnetized plasmas},
  journal = {Phys. Rev. E},
  volume = {82},
  number = {5},
  pages = {056407},
  year = {2010},
  doi = {10.1103/PhysRevE.82.056407},
}

@article{zhang2020relativistic,
  author = {Zhang, P and Bulanov, S. S. and Seipt, D and Arefiev, A. V. and Thomas, A. G. R.},
  title = {Relativistic plasma physics in supercritical fields},
  journal = {Phys. Plasmas},
  volume = {27},
  number = {5},
  pages = {050601},
  year = {2020},
  doi = {10.1063/1.5144449},
}
\end{document}